\documentclass[twocolumn]{aastex62}

\usepackage{natbib} 
\usepackage{hyperref}
\usepackage{xcolor}
\usepackage{amsmath}
\usepackage{rotating}
\usepackage{array}
\usepackage{bibentry}
\usepackage{amsmath}
\usepackage{enumitem} 
\usepackage{booktabs}
\usepackage{multirow}
\usepackage{gensymb}
\newcolumntype{H}{>{\setbox0=\hbox\bgroup}c<{\egroup}@{}}

\hypersetup{colorlinks,
linkcolor={red!50!black},
	citecolor={blue!60!black},
	urlcolor={pink!40!black}
}

\newcommand{\hiz}{high-$z$}

\newcommand{\mum}{$\mu$m }
\newcommand{\spit}{\emph{Spitzer}}
\newcommand{\her}{\emph{Herschel}}
\newcommand{\vm}{$v_{\rm max}$}
\newcommand{\msun}{$M_{\odot}$}

\newcommand{\my}{$M_{\odot}$ yr$^{-1}$}

\newcommand{\kms}{km s$^{-1}$}
\newcommand{\mout}{$\dot M_{\rm OF}$}

\newcommand{\aco}{$\alpha_{\rm CO}$}
\newcommand{\acou}{$M_{\odot} $(K km s$^{-1}$ pc$^{-2}$)$^{-1}$}
\newcommand{\lu}{K km s$^{-1}$ pc$^{-2}$}
\shorttitle{}
\shortauthors{Gowardhan et al}

\begin{document}

\def\andname{\hspace*{-0.5em}}
	
\author[0000-0002-3310-5859]{Avani Gowardhan} 
\affiliation{Department of Astronomy, Cornell University, Ithaca, NY 14853, USA}

\author{Henrik Spoon}
\affiliation{Department of Astronomy, Cornell University, Ithaca, NY 14853, USA}

\author{Dominik A. Riechers}
\affiliation{Department of Astronomy, Cornell University, Ithaca, NY 14853, USA} 

\author{Eduardo Gonz\'{a}lez-Alfonso}
\affiliation{Universidad de Alcal\'{a}, Departamento de F'{i}sica y Matem\'{a}ticas, Campus Universitario, E-28871 Alcal\'{a} de Henares, Madrid, Spain}
\affiliation{Harvard-Smithsonian Center for Astrophysics, 60 Garden Street, Cambridge, MA 02138, USA}

 \author{Duncan Farrah}
 \affiliation{Department of Physics, Virginia Tech, Blacksburg, VA 24061, USA } 

\author{Jacqueline Fischer}
\affiliation{Naval Research Laboratory, Remote Sensing Division, 4555 Overlook Avenue SW, Washington DC 20375, USA }

\author{Jeremy Darling}
\affiliation{Center for Astrophysics and Space Astronomy, Department of Astrophysical and Planetary Sciences, University of Colorado, 389 UCB, Boulder, CO 80309-0389, USA} 

\author{Chiara Fergulio}
\affiliation{INAF Osservatorio Astronomico di Trieste, via G.B. Tiepolo 11, 34143, Trieste, Italy} 

\author{Jose Afonso}

\affiliation{Instituto de Astrof\'{i}sica e Ci\^{e}ncias do Espa\c co, Universidade de Lisboa, OAL, Tapada da Ajuda, PT1349-018 Lisboa, Portugal}

\affiliation{Departamento de F\'{i}sica, Faculdade de Ci\^{e}ncias, Universidade de Lisboa, Edif\'{i}cio C8, Campo Grande, PT1749-016 Lisbon, Portugal}

\author{Luca Bizzocchi}

\affiliation{Center for Astrochemical Studies, Max-Planck-Institut f\"{u}r extraterrestrische Physik, Gie{\ss}enbachstra{\ss}e 1, 85748 Garching, Germany}

\correspondingauthor{Avani Gowardhan}
\email{ag2255@cornell.edu}

\title{The dual role of starburst and active galactic nuclei in driving extreme molecular outflows}

\begin{abstract}

We report molecular gas observations of IRAS 20100$-$4156 and IRAS 03158$+$4227, two local ultraluminous infrared galaxies (ULIRGs) hosting some of the fastest and most massive molecular outflows known. Using ALMA and PdBI observations, we spatially resolve the CO($1-0$) emission from the outflowing molecular gas in both and find maximum outflow velocities of \vm$\sim 1600$ and $\sim 1700$ \kms$ $ for IRAS 20100$-$4156 and IRAS 03158$+$4227, respectively. We find total gas mass outflow rates of $\dot M_{\rm OF}\sim 670$ and $\sim 350$ \my, respectively, corresponding to molecular gas depletion timescales $\tau^{\rm dep}_{\rm OF} \sim 11$ and $\sim 16$ Myr. This is nearly $3$ times shorter than the depletion timescales implied by star formation, $\tau^{\rm dep}_{\rm SFR}\sim 33$ and $\sim 46$ Myr, respectively. To determine the outflow driving mechanism, we compare the starburst ($L_{*}$) and AGN ($L_{\rm AGN}$) luminosities to the outflowing energy and momentum fluxes, using mid-infrared spectral decomposition to discern $L_{\rm AGN}$. Comparison to other molecular outflows in ULIRGs reveals that outflow properties correlate similarly with $L_{*}$ and $L_{\rm IR}$ as with $L_{\rm AGN}$, indicating that AGN luminosity alone may not be a good tracer of feedback strength and that a combination of AGN and starburst activity may be driving the most powerful molecular outflows. We also detect the OH 1.667 GHz maser line from both sources and demonstrate its utility in detecting molecular outflows.
\end{abstract}
\section{Introduction} \label{sec:intro}

Numerous observations suggest a close evolutionary relationship between supermassive black hole (SMBH) growth and stellar assembly in massive galaxies. These include the strong correlation between SMBH mass ($M_{\rm BH}$) and stellar bulge properties i.e. its mass ($M_{\rm bulge}$) and velocity dispersion ($\sigma$; \citealt{magorrian1998, ferrarese2000, gebhardt2000, mcconnell2013}). Furthermore, the observed galaxy luminosity function declines more steeply at high luminosities $L > L^{*}$ \citep{baugh1998} than predicted based on $\Lambda$CDM hierarchical structure formation models, suggesting that star formation is suppressed in massive galaxies ($M^{*}> 10^{11}$\msun$ $ at $z\sim 0$). Galaxy evolution models often invoke some form of feedback from active galactic nuclei (AGN) to quench star formation and reconcile luminosity function models with these observations \citep{benson2003,bower2006,somerville2008}. Two main forms of feedback have been incorporated into models - radio (`maintenance mode') and radiative (`quasar mode') feedback. Radio mode feedback is associated with lower accretion rates \citep[e.g.][]{ishibashi2014b}, takes place in the form of highly-collimated and relativistic radio jets, and is typically seen in early-type galaxies (i.e. in giant ellipticals at low-$z$). In contrast, radiative feedback results when the SMBH approaches the Eddington accretion rate, producing energetic nuclear winds that collide with the interstellar medium (ISM), generating a blast wave which drives out the dust and cold gas in the form of massive outflows, thus quenching star formation and growth of the stellar bulge \citep{lapi2005,king2010,zubovas2012}. In the classic galaxy evolution picture, this is the crucial step in the evolution from dusty, obscured AGN host galaxies to unobscured and optically luminous quasars, and therefore important at \hiz. Local ultraluminous infrared galaxies (ULIRGs) hosting powerful AGN and/or starburst activity are ideal targets in which to study feedback physics and its role in galaxy evolution, being local analogs of the dusty star-forming galaxy (DSFG) population at \hiz$ $ \citep[see][for a review]{casey2014}. 

While ULIRGs often show prominent ionized and neutral gas outflows \citep[e.g.][]{rupke2005,spoon2009}, observations of outflowing molecular gas - the fuel for star formation - are necessary to definitively identify quenching of star formation. Molecular outflows have been detected in \emph{Herschel} observations of ULIRGs using far-IR transitions (OH 119 \mum and OH 79 $\mu$m), which trace the outflows close to the central nucleus ($\sim 300$ pc, flow times $\tau \sim $ few$\times 10^{5}$ yrs; \citealt{fischer2010, sturm2011, ga2017}) and show a characteristic P-Cygni profile, providing unambiguous evidence of outflowing gas. Further observations of local ULIRGs demonstrate that energetically dominant AGN hosts have higher maximum gas outflow velocities (up to $\sim 1400-2000$ \kms; \citealt{fischer2010, sturm2011,spoon2013,veilleux2013}) as compared to those dominated by starburst activity. CO($1-0$) observations of such ULIRGs trace the outflow on larger spatial (up to $\sim 10$ kpc) and longer time scales (flow times $\tau \gtrsim 10^{6}$ yrs) than the far-IR OH transitions, and demonstrate that the molecular gas can be depleted on short timescales $\sim 10$ Myr \citep[e.g][]{feruglio2010,cicone2014}. Finally, both OH and CO($1-0$) observations for a sample of nearby galaxies hosting outflows show increasing mass outflow rates and higher outflow velocities with increasing AGN luminosities, interpreted by the authors as more powerful molecular outflows being mostly powered by AGN \citep[e.g.][]{fischer2010, sturm2011, spoon2013,veilleux2013, cicone2014,stone2016}

However, there are many open questions in this picture of galaxy evolution. While outflows in massive galaxies are ubiquitous, their impact on star formation is still debated, with some studies finding that outflows decrease the star formation \citep[e.g.][]{farrah2012, alatalo2015a}, and others finding no difference in star formation \citep[e.g][]{violino2016,pitchford2016, balmaverde2016}. The driving mechanism of outflows is also still an open question. Nuclear starburst and AGN activity are often observed concurrently in ULIRGs \citep[e.g.][]{genzel1998, farrah2002, farrah2003,veilleux2009, efstathiou2014, xu2015, rosenberg2015}. This makes it difficult to identify the respective roles of AGN and starburst activity in ULIRGs, especially given the high nuclear dust obscuration in most ULIRGs. It is also debated how nuclear starbursts relate to AGN activity: numerical simulations variously show that starbursts can both precede AGN activity and SMBH growth \citep[e.g.][]{hopkins2012}, or can be triggered by AGN feedback \citep[e.g.][]{silk2010, ishibashi2013}. It has also been proposed that both star formation and SMBH growth are independent and self-regulatory \citep{cen2012} and that the observed bulge-BH correlations are produced by hierarchical structure formation \citep[e.g.][]{jahnke2011} rather than feedback processes. Observations of outflowing molecular gas provide critical insight into these open questions. Molecular gas not only traces the fuel for star formation but can also make up the bulk of the outflowing gas mass. Spatially resolved molecular gas observations can be used to determine the outflow energetics, distinguish between outflow models, and determine the driving mechanism. 

Here we present observations of two local ULIRGs hosting the most extreme molecular outflows in the local Universe - IRAS 20100$-$4156 and IRAS 03158$+$4227, both at $z\sim0.13$. The sources have been selected from the \emph{Herschel} ULIRG Survey (HERUS), a complete, flux-limited ($S_{60 \mu \rm m} >1.9$ Jy) far-IR spectroscopic sample of the local ULIRG population \citep{farrah2013,herus}. While molecular outflows are ubiquitous in this and other samples of local ULIRGs \citep{spoon2013,veilleux2013}, the two selected sources display the deepest absorption profiles in far-IR OH lines and/or the highest maximum outflow velocities in their OH 119 \mum spectra \citep{spoon2013, ga2017}, even more so than the archetypal outflow source, Mrk 231 \citep{fischer2010, spoon2013, ga2017}. Such high-velocity outflows are expected to be much more common at \hiz$ $ \citep[e.g.][]{maiolino2012}, making it crucial to understand their origin and impact on the host galaxy. We present spatially resolved observations of the outflowing molecular gas in both these sources - traced using CO($1-0$) - to determine the outflow properties (geometry, outflow rate, and energetics), while simultaneously using multi-wavelength spectroscopy and extensive photometric coverage (from mid-IR to X-ray) to identify the relative AGN and starburst contributions, and relate them to the outflow properties. Finally, we also present observations of the OH 1.667 GHz maser line (hereafter OHM) in our targets and demonstrate its utility for future molecular gas studies. 

The paper is organized as follows: in \autoref{sec:obs} and \autoref{sec:results}, we present the observations and results. In \autoref{sec:analysis}, we discuss the outflow properties, spectral energy distribution (SED) fitting, and mid-IR spectral decomposition. In \autoref{sec:discussion} and \autoref{sec:conclusions}, we discuss our results and conclusions. We use a $\Lambda$CDM cosmology, with $H_{0} = 71$ km s$^{-1}$ Mpc$^{-1}$, $\Omega_{\rm M} = 0.27$, and $\Omega_{\Lambda} = 0.73$ \citep{spergel2007}.

\section{Observations}\label{sec:obs}

In this section, we discuss the details of molecular gas observations (summarized in \autoref{tab:obs}) as well as existing observations in the literature in \autoref{ssec:arch}. 

\subsection{CO observations}\label{ssec:co}
\subsubsection{CO($1-0$) in IRAS 20100$-$4156}\label{sssec:iras20100_co10_obs}

We observed the CO($1-0$) line in IRAS 20100$-$4156 using the Atacama Large Millimeter Array (ALMA), over 5 epochs in June 2014 with 37 antennas (PI: Spoon). The absolute flux scale was calibrated using Neptune and Uranus, while J$2056-4714$ was used as a bandpass and phase calibrator. The receivers were tuned to a central frequency of 102.055 GHz, with a bandwidth of 1.875 GHz in each of 4 spectral windows, corresponding to 3840 frequency channels with a channel width of 0.49 MHz ($\sim 1.4$ \kms)

CASA $v4.2$ was used for the ALMA pipeline calibration, and the resulting measurement set was used for all further analysis. Continuum subtraction was performed using the task {\sc uvcontsub}, using line-free channels and spectral windows (a total bandwidth of 2902 MHz), using a $1^{\rm st}$ order polynomial baseline for the continuum. This calibrated and continuum-subtracted visibility dataset was used for the task {\sc uvmodelfit}. Prior to using {\sc uvmodelfit}, the task {\sc statwt} was used to calculate the weights for the visibilities based on line-free channels only. The reduced measurement set was imaged and cleaned using the CASA task {\sc clean}, using natural weighing and a pixel size of $0''.3 \times 0''.3$, after binning over 71 spectral channels (corresponding to $\sim 100 $ \kms). The resulting cleaned image has an rms noise of 0.1 mJy beam$^{-1}$ over each channel with width 34 MHz, (corresponding to $\sim 100.2$ \kms$ $ at 102.055 GHz), and a resulting synthesized beam size of $1''.5 \times 1''.2$ (PA = 75\degree). As the emission is resolved both in velocity as well as spatially, the clean mask for the emission region was selected after visual inspection for each channel. 
 
\subsubsection{CO($3-2$) in IRAS 20100$-$4156}\label{sssec:iras20100_co32_obs}

We observed CO($3-2$) emission in IRAS 20100$-$4156, redshifted to $\nu_{\rm obs} = 306.082$ GHz, using the single dish sub-mm telescope Atacama Pathfinder Experimental telescope (APEX) over 8 epochs from August - October 2012 (PI: Farrah), for a total on-source time of 3.7 hours. The weather ranged from good (with the atmospheric precipitable water vapor (pwv) $\sim 0.5$ mm) to acceptable (pwv $\sim 4$mm) during the observations. The APEX-2 heterodyne receiver was used as a front-end with the eXtended bandwidth Fast Fourier Transform Spectrometer (XFFTS) as the backend, with wobbler-switching at a frequency of 0.5 Hz and an asymmetrical azimuthal throw of $50''$. Saturn and RZ-Sgr were used for pointing. The total bandwidth was 2.5 GHz, centered at 306.0119 GHz, with a frequency resolution of 76.3 kHz, which corresponds to a velocity resolution of 0.074 km s$^{-1}$. The half-power beam width at this frequency is $\sim 21''$, so the emission was unresolved. Data reduction was carried out using GILDAS {\sc class} (Continuum and Line Analysis Single-dish Software), and the spectrum was binned to a velocity resolution of $\sim 53$ \kms. The antenna temperatures ($T_{\rm A}$) were converted to flux densities assuming an aperture efficiency of $\eta_{\rm a}=0.6$, which implies a K to Jy conversion factor $24.4/$ $\eta_{\rm a} = 41$ Jy K$^{-1}$ \footnote{\url{http://www.apex-telescope.org/telescope/efficiency}}. The final spectrum has an rms noise of 0.6 mJy over a line-free bandwidth of $\sim 3300$ \kms. 

\subsubsection{CO($1-0$) in IRAS 03158$+$4227}\label{sssec:iras03158_co10_obs}

We observed CO($1-0$) emission toward IRAS 03158$+$4227 with the Plateau de Bure Interferometer (PdBI) in the most compact 5 antenna configuration (5D), during July-August 2013 (PI: Spoon). The weather was average during these observations (pwv $\sim10$ mm on average). The quasars 3C84 and B0234+285 were used as phase and amplitude calibrators respectively, and the quasars 3C84 and 3C345 were used as RF bandpass calibrators. The star MWC349 and the quasar 3C454.3 were used as absolute flux calibrators. The WideX correlator with a bandwidth of 3.6 GHz was tuned to 101.604 GHz, and the observations were carried out in dual polarization mode, with a spectral resolution of 1.95 MHz (corresponding to $\sim 5.8$ km s$^{-1}$). The total on-source time, after flagging of bad visibilities was 19.5 hours (6 antenna equivalent time, summing together 18 observations). Given the large bandwidth of the observations, we also covered CN($N= 1 - 0$, $J=3/2-1/2$), redshifted from $\nu_{\rm rest} \sim 113.490$ GHz to $\nu_{\rm obs} = 100.04612$ GHz. 

All observations were calibrated using the IRAM PdBI data reduction pipeline in {\sc clic} (Continuum and Line Interferometer Calibration), with subsequent additional flagging by hand. The absolute flux scale was calibrated to better than $20\%$. Continuum subtraction was performed for the reduced visibility cube using the task {\sc uv\_subtract}, excluding the line channels (a total of 2.64 GHz of line-free bandwidth). The resulting continuum-subtracted visibility data was imaged using {\sc uv\_map}, with natural weighing and a pixel size of $0''.5 \times 0''.5$, and cleaned using the task {\sc clean} with the Hogbom algorithm at a binned spectral resolution of 20 MHz ($\sim 60$ \kms$ $ at a frequency of $101.604$ GHz). The resulting cleaned image has an rms noise of 0.26 mJy beam$^{-1}$ over each 20 MHz channel, and a beam size of $4''.8 \times 3''.9$ (PA = 166\degr). Primary beam correction was applied using the task {\sc primary}. 

\subsection{OHM observations}\label{ssec:ohm_obs}

We observed the OH 1.667 GHz maser line for both IRAS 20100$-$4156 and IRAS 03158$+$4227, using the NSF's Karl G. Jansky Very Large Array (VLA) in August 2015, as part of a Director's Discretionary Time (DDT) proposal (PI: Spoon). Observations were made using the L band receivers in the most extended configuration (A array), using 27 antennas. We used a bandwidth of 1 GHz over 8 spectral windows with a bandwidth of 128 MHz for each sub-band, with spectral resolutions of 111 kHz and 142 kHz in the spectral windows covering the lines for IRAS 20100$-$4156 and IRAS 03158$+$4227, respectively. The receivers were tuned to central frequencies of 1.411 GHz and 1.407 GHz, respectively. 

IRAS 20100$-$4156 was observed for a total on-source time of 90 minutes. 3C48 and J1937$-$3958 were used for absolute flux and phase/bandpass calibration respectively. IRAS 03158$+$4227 was observed for a total on-source time of 20 minutes. 3C147 and J0319+4130 were used for absolute flux and phase/bandpass calibration, respectively. 

The VLA reduction pipeline in CASA $v 4.3.1$ was used to flag and calibrate the observations. Continuum subtraction was performed based on the line-free channels (total line-free bandwidth of $\sim 0.7$ GHz for both using the task {\sc uvcontsub}, using a $1^{\rm st}$ order fit for the continuum. For IRAS 20100$-$4156, the reduced measurement set was cleaned using the CASA task {\sc clean}, using natural weighting and a pixel size of $0.''5 \times 0.''5$. The resulting cleaned image has an rms noise of 0.8 mJy beam$^{-1}$ over each channel of 142 kHz (corresponding to $\sim 30.2 $ \kms$ $ at 1.411 GHz), and a synthesized beam size of $4''.9 \times 1''.0$ (PA: $-175\degree$). 

For IRAS 03158$+$4227, the same pipeline was used for flagging and calibration. Continuum subtraction was performed using a linear $1^{\rm st}$ order fit to line-free channels. The CASA task {\sc clean} was used with natural weighting and a pixel size of $0''.5 \times 0''.5$, after binning over 5 velocity channels (each corresponding to 111 kHz). The resulting cleaned image has an rms noise of 1.0 mJy beam$^{-1}$ over each 565 kHz channel (corresponding to a velocity resolution of 120 \kms$ $ at 1.407 GHz) and a synthesized beam size of $1''.6 \times 1''.3$ (PA: $-45.58$\degree).

%% table to describe observations 
\begin{table*}
\begin{center}
\caption{\added{Details of} molecular gas observations of IRAS 20100$-$4156 and IRAS 03158+4227.\explain{The spectroscopic notation for the OH line has been corrected, and the CN line has been added.}}
\begin{tabular}{cccllHll}
\tableline

Source		& Position 	& $z_{\rm CO}$ 	& Telescope &  Transition  &  $\nu_{\rm rest}$ & Beam size & $T_{\rm on}$ (hours)   \\
\tableline
\tableline
IRAS 20100$-$4156		& $20^h13^m29.5^s -41^d47^m35^s$ 	&  $0.12975$ 	&  ALMA  				& CO($1-0$)	& 115.27120 &   $1''.5 \times 1''.2 $		&	4.1 \\
					&								&			&  VLA 				&  OH($^{2}\Pi_{3/2}, J = 3/2, F = 2^{e} - 2^{f}$) & 1.66736 &  $4''.9 \times 1''.0 $	 	& 	1.5 \\
					&								&			&  APEX 				& CO($3-2$)	&  345.79599 &  $21'' $	  				&	3.7 		\\
 IRAS 03158+4227		&$03^h19^m12.4^s +42^d38^m28^s$ 	& $ 0.13459$	& PdBI 				& CO($1-0$) & 115.27120 &  $4''.8 \times 3''.9 $ &19.5 	\\
& & & PdBI & CN ($N = 1 - 0, J= 3/2 - 1/2$)$^{a}$ & 101.11 &  $4''.8 \times 3''.9 $ & 19.5 \\ 
					&								&			& VLA 				& OH($^{2}\Pi_{3/2}, J = 3/2, F = 2^{e} - 2^{f}$)  & 1.66736	&   $1''.6 \times 1''.3$ 			& 0.3 		\\
                    
\tableline

\end{tabular}
\label{tab:obs}
\end{center}
\tablenotetext{a}{\added{We observed only one component of the CN doublet.}}
\end{table*}

\subsection{Archival Data}\label{ssec:arch}

Extensive photometric and spectroscopic coverage exists for both our target sources. The photometric coverage includes X-ray observations from XMM Newton, far-UV and near-UV observations from GALEX, and both mid- and far-IR observations (spectroscopy as well as photometry) using \emph{Herschel} PACS/SPIRE and the \spit$ $ InfraRed Spectrograph (IRS; \citealt{houck2004}). Details of the photometry used and their references are listed in \autoref{tab:phot}. Optical imaging was obtained using the \emph{Hubble Space Telescope} Wide Field and Planetary Camera 2 (HST/WFPC2) instrument (F814W band) for IRAS 20100$-$4156 (as part of the proposal IDs 6346, 7896, \citealt{cui2001, bushouse2002}), and the Sloan Digital Sky Survey (SDSS) for IRAS 03158$+$4227; \citep{sdss2016}. Astrometric corrections were determined for the F814W HST image using nearby stellar positions from the General Star Catalog 2, leaving residual uncertainties of $0''.2$.\emph{Herschel} PACS observations of the far-IR fine-structure lines and OH 119 \mum and 79 \mum profiles were presented by \citet{farrah2013} and \citet{spoon2013}, and modeling of these far-IR OH lines (as well as the OH 84 \mum and OH 65 \mum doublets) was presented in \cite{ga2017}. 

We compare the outflow properties of IRAS 20100$-$4156 and IRAS 03158$+$4227 against molecular outflows in other local ULIRGs with published OH 119 \mum and CO observations \citep{cicone2014, garcia2015,veilleux2017} in order to place our results in the larger context of molecular outflows using the most uniform sample possible based on outflow and galaxy properties. We use \spit$ $ mid-IR spectra for all the sources, available from the Combined Atlas of Sources with \emph{Spitzer} IRS Spectra (CASSIS; \citealt{lebouteiller2011,lebouteiller2015}). The high sensitivity and spectral resolution of CO($1-0$) observations mean that the CO-based redshifts are more accurate than previously accessible, and we use those for the rest of the paper. 

\setlength{\tabcolsep}{1.0pt}
\renewcommand{\arraystretch}{0.8} 
\begin{deluxetable}{lccccc} \tablewidth{0pt}
\tablecaption{Photometry for IRAS 20100$-$4156 and IRAS 03158$+$4227.}
\tablehead{
\colhead{Telescope}           & \colhead{Band}      &
\colhead{$\lambda_{\rm eff}$($\mu$m)}& \colhead{Aperture$^a$}  &
\colhead{Flux (mJy)}          & \colhead{ Ref}}
\startdata
\vspace{-2 mm} \\ 
IRAS 20100$-$4156 & & & &  \\ 
\vspace{-2 mm } \\ 
\hline 
\vspace{-2 mm } \\ 
GALEX 	& FUV		& 0.15	& $7''.5$	& 	$(3.0 \pm 0.4) \times 10^{-2} $	& (1) \\
	 			& NUV 		& 0.23	& $7''.5$	& 	$(6.8 \pm 0.4) \times 10^{-2} $   	& (1) \\
		$HST$ 	& F814W 		& 0.8 	& PF		& 	$(1.3 \pm 0.1) $ 							& (2) \\ 
		2MASS    & J	 		& 1.24	& $20''$	&  	$(2.3 \pm 0.2)$     	& (3) \\ 
	 			& H			& 1.65	& $20''$	&  	$(1.9 \pm 0.4)$  	& (3) \\ 
	 			& Ks			& 2.17	& $20''$	&  	$(2.8 \pm 0.4) $ 	& (3) \\ 
		WISE	& W3		& 12		& $8''.3$	& 	$(4.1 \pm 0.1) \times 10^{1}$		& (4) \\
	\emph{Spitzer}	& IRS  		& 11		& $3''.7^b$	&  	$(6.7 \pm 0.4) \times 10^{1}$  		& (5) \\
	 			& IRS  		& 13.5	& $3''.7^b$	&  	$(3.1 \pm 0.1) \times 10^{1}$  	& (5) \\
	 			& IRS  		& 17		& $10''.5^b$	&  	$(1.0 \pm 0.1) \times 10^{2}$  	& (5) \\
	 			& IRS  		& 23		& $10''.5^b$	&  	$(4.1 \pm 0.1) \times 10^{2}$  		& (5) \\
				& MIPS 		& 24		&  PF 		&  	$(2.8 \pm 0.1) \times 10^{2}$ 	& (6) \\
	 			& IRS 		& 26		&  $10''.5^b$	&  	$(8.3 \pm 0.1) \times 10^{2}$  	& (5) \\
	 			& IRS  		& 29		&  $10''.5^b$	&  	$(1.3 \pm 0.1) \times 10^{3}$  	& (5) \\
  			\her  	& cont.		& 64    	& $45''$   	&   	$(5.8 \pm 0.6) \times 10^{3}$ 		& (7)\\  
		   		& cont. 		& 78    	& $45''$  	&   	$(5.3 \pm 0.5) \times 10^{3}$ 		& (7)\\  
	  			& cont.		& 84    	& $45''$   	&   	$(5.2 \pm 0.5) \times 10^{3}$   	& (7)\\  
	 			& cont.		& 120   	& $45''$  	&  	$(3.9 \pm 0.4) \times 10^{3}$    	& (7)\\  
	   			& cont.		& 131   	& $45''$   	&   	$(3.5 \pm 0.4) \times 10^{3}$   	& (7)\\  
	 			& cont.		& 146   	& $45''$   	&   	$(2.7 \pm 0.3) \times 10^{3}$   	& (7)\\  
				& cont. 		& 158   	& $45''$   	& 	$(2.4 \pm 0.2) \times 10^{3}$    	& (7)\\  
				& cont. 		& 169   	& $45''$   	&  	$(1.9 \pm 0.2) \times 10^{3}$ 	& (7)\\  
				& SPIRE		& 250   	& $17''.6$ 	&	$(1.0 \pm 0.1) \times 10^{3}$	& (8)\\  
				& SPIRE 		& 350   	& $23''.9$	&   	$(3.5 \pm 0.1) \times 10^{2}$	& (8)\\  
				& SPIRE		& 500   	& $35''.2$	&   	$(1.2 \pm 0.3) \times 10^{2}$	& (8)\\ 	
	ALMA		& cont. 		& $3.0\times 10^{3}$	&  	PF	& ($1.6 \pm 0.3$)		& (2)\\ 
	VLA 			& L band		& $2.1\times 10^{5}$	& 	PF 	&  $19.6 \pm 4.0$ 	  	& (2)\\ 
\hline 
\vspace{-2 mm} \\ 
IRAS 03158$+$4227 & & & &  \\ 
\vspace{-2 mm } \\ 
\hline 
\vspace{-2 mm } \\ 
    	SDSS	& 	$u'$ 	  	& 0.35	& PF		& 	$ (1.9 \pm 0.2) \times 10^{-2}$	 	& (9)\\
			& 	$g'$ 	   	& 0.48  	& PF 	&   	$ (1.2 \pm 0.1) \times 10^{-1}$  		& (9)\\
			&	$r'$ 		& 0.62  	& PF		&  	$ (2.8 \pm 0.3) \times 10^{-1}$  		& (9)\\
			& 	$i'$ 		& 0.76  	& PF		&  	$ (4.5 \pm 0.5) \times 10^{-1}$  		& (9)\\
			& 	$z'$ 	   	& 0.91 	& PF 	&   	$ (5.5 \pm 0.6) \times 10^{-1}$  		& (9)\\	
	\emph{Spitzer}	
			& IRS 		& 11 		& $3''.7^b$ 		& $(4.3 \pm 0.9$)  				& (5)\\
			& IRS 		& 13.5 	& $3''.7^b$ 		& $(4.4 \pm 0.2 \times 10^{1}$  		& (5)\\
			& IRS 		& 17 		& $10''.5^b$ 	& $(1.5 \pm 0.1) \times 10^{2}$  	& (5)\\
			& IRS 		& 23 		& $10''.5^b$	& $(4.9 \pm 0.1) \times 10^{2}$ 	& (5)\\
			& IRS 		& 26 		& $10''.5^b$ 	& $(9.0 \pm 0.1) \times 10^{2}$  	& (5)\\ 
			& IRS 		& 29 		& $10''.5^b$ 	& $(1.4 \pm 0.1) \times 10^{3}$ 	& (5)\\ 
   	\her   	& cont.		& 64		&$45''$ 		& $(4.9 \pm 0.6) \times 10^{3}$ 	& (7)\\  
	   		& cont. 		& 78    	&$45''$ 		& $(4.9 \pm 0.5) \times 10^{3}$ 	& (7)\\  
	  		& cont.		& 84    	&$45''$ 		& $(4.8 \pm 0.5) \times 10^{3}$ 	& (7)\\  
	 		& cont.		& 120  	&$45''$		& $(3.1 \pm 0.4) \times 10^{3}$ 	& (7)\\  
	   		& cont.		& 131   	&$45''$ 		& $(2.9 \pm 0.4) \times 10^{3}$ 	& (7)\\  
	 		& cont.		& 146  	&$45''$ 		& $(2.7 \pm 0.3) \times 10^{3}$ 	& (7)\\  
			& cont. 		& 158  	&$45''$ 		& $(2.2 \pm 0.2) \times 10^{3}$ 	& (7)\\  
			& cont. 		& 169   	&$45''$ 		& $(2.1 \pm 0.2) \times 10^{3}$ 	& (7)\\  
			& SPIRE		& 250   	&$17''.6$		& $(9.7 \pm 0.1) \times 10^{3}$ 	& (8)\\  
			& SPIRE 		& 350   	&$23''.9$		& $(3.8 \pm 0.1) \times 10^{2}$ 	& (8)\\  
			& SPIRE		& 500  	 &$35''.2$		& $(1.4 \pm 0.3) \times 10^{2}$ 	& (8)\\ 	
	NOEMA	& cont. 		& $3.0 \times 10^{3}$ & PF &  $(9.5 \pm 0.1) \times 10^{-1}$	& (2)\\ 
	VLA 		& L band		& $2.1 \times 10^{5}$ & PF &  $(1.6 \pm 0.1) \times 10^{1}$	& (2) \\ 
\enddata
\tablenotetext{a}{Aperture for flux-extraction; PF stands for point-source fitting and extraction.}
\tablenotetext{b}{Slit width for IRS spectra.}
\tablerefs{(1) GALEX Data Release 6 Catalog (2) This work (3) 2MASS All Sky Survey, \citet{skrutskie2006} (4) WISE All Sky Data Release (5) CASSIS, \citet{lebouteiller2011,lebouteiller2015} (6) \emph{Spitzer},  \citet{capak2013} (7) \citealt{ga2017} (8) \citet{pearson2016}(9) SDSS DR IV, \citet{blanton2017}.}
\label{tab:phot}
\end{deluxetable}

\section{Results}\label{sec:results}
\subsection{IRAS 20100$-$4156} \label{ssec:iras20100}

%%%%%%%%%%%%%%%%
%figures 
%%%%%%%%%%%%%%%%

\begin{figure*}
	\centering
	\includegraphics[width=\textwidth]{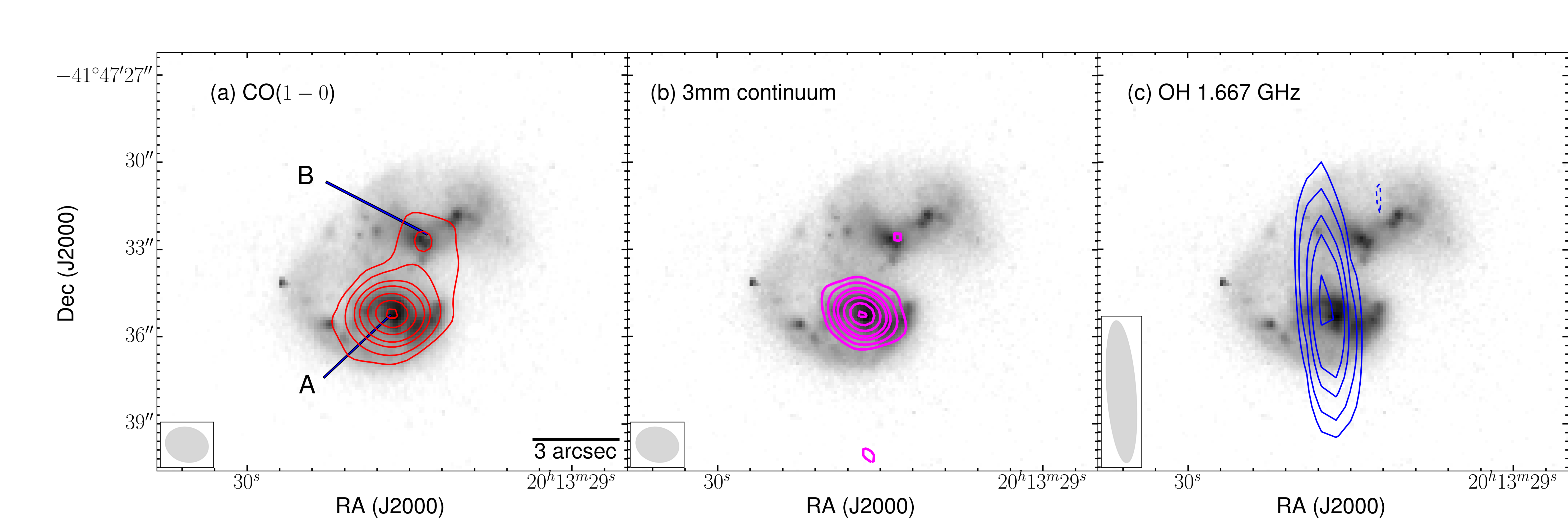}
	\caption{(a) Moment-0 map of CO($1-0$) emission from IRAS 20100$-$4156, created by integrating over velocity range $(-1169, 1650)$ \kms, overlaid on a HST/WFPC2 F814W image (b) 3mm continuum emission (c) Moment-0 map of the OHM emission, created by integrating over the velocity range $(-927,527)$ \kms, the peak of which is spatially offset from the peak of the dust and CO($1-0$) emission (see \autoref{sssec:iras20100_ohm}). Each panel shows contours marked at $\pm 3,7,15,23,40,60,80\sigma$ significance levels. The peaks of the two components of the merger (marked A and B) show a projected separation of $5.6$ kpc. We detect the molecular gas outflow from component A. Synthesized beam sizes are $1''.5\times 1''.2$ for (a) and (b), and $4''.9 \times 1''.0$ for (c), and are shown at the bottom left of each panel. } 
	\label{fig:iras20100_hst_cont_oh}
\end{figure*}

\begin{figure}
	\centering
	\includegraphics[width=0.45\textwidth]{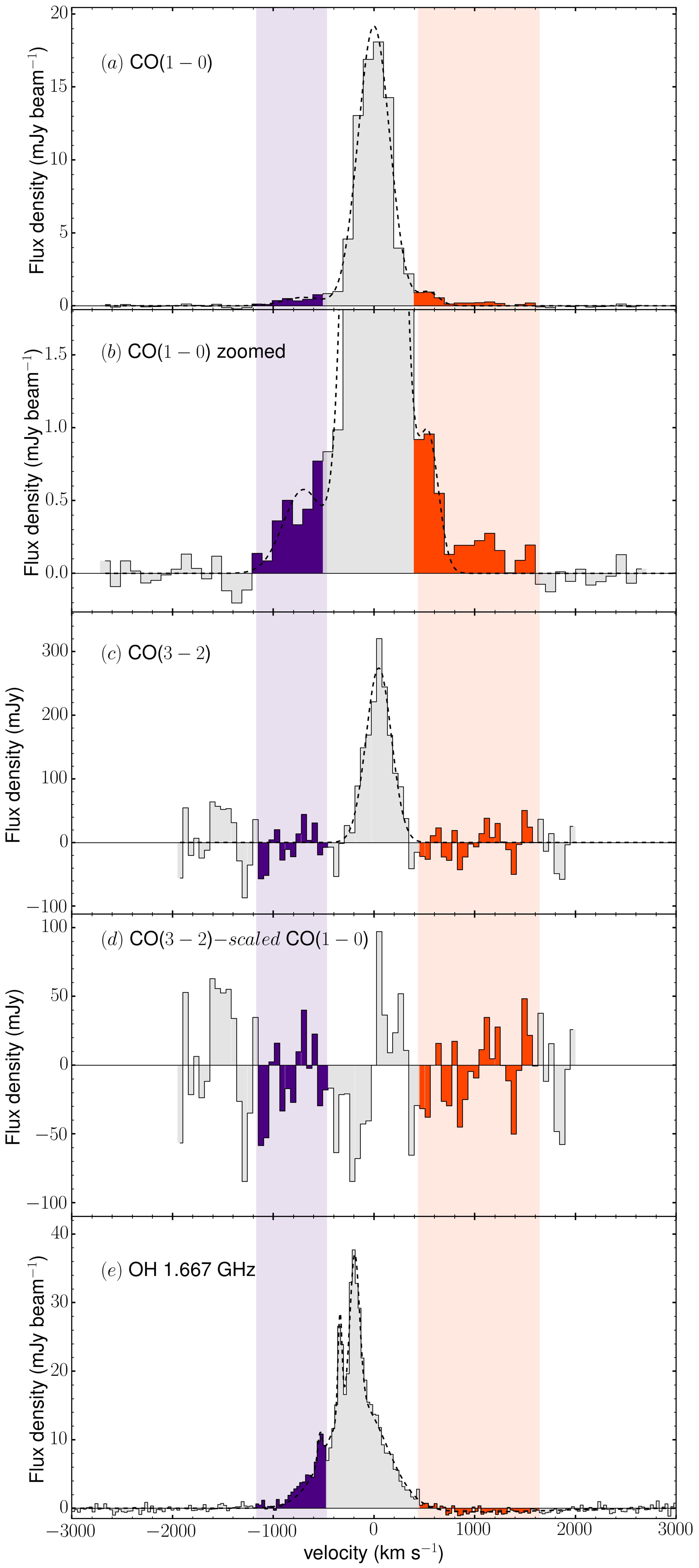}
	\caption{(a) CO($1-0$) spectrum extracted from the peak pixel of component A in IRAS 20100$-$4156 (b) Same as (a), zoomed in to show the high-velocity outflow wings (c) CO($3-2$) emission from IRAS 20100$-$4156 (d) Difference between CO($3-2$) and scaled CO($1-0$) line profiles. The CO($1-0$) line profile has been binned and re-normalized to have the same spectral resolution and the same flux at $v \sim 0$ \kms$ $ as CO($3-2$) (e) OHM spectrum for IRAS 20100$-$4156 extracted from the peak of OHM emission. The dashed lines show the sum of best fit Gaussians, using a combination of three Gaussians for (a) and (b), and four Gaussians for (e). The purple and orange regions show the velocity ranges for the blue and red outflow wings defined for the CO($1-0$) spectrum (see \autoref{sssec:iras20100_wings}), and $v = 0$ \kms$ $ has been defined using $z_{\rm CO}$. A systemic offset of $\Delta v \sim 190$ \kms$ $ between the peaks of the CO($1-0$) and OHM line emission can be seen upon comparing (a) and (e).}
\label{fig:iras20100_allspecs}
\end{figure}

\begin{figure*}
	\centering
	\includegraphics[width=\textwidth]{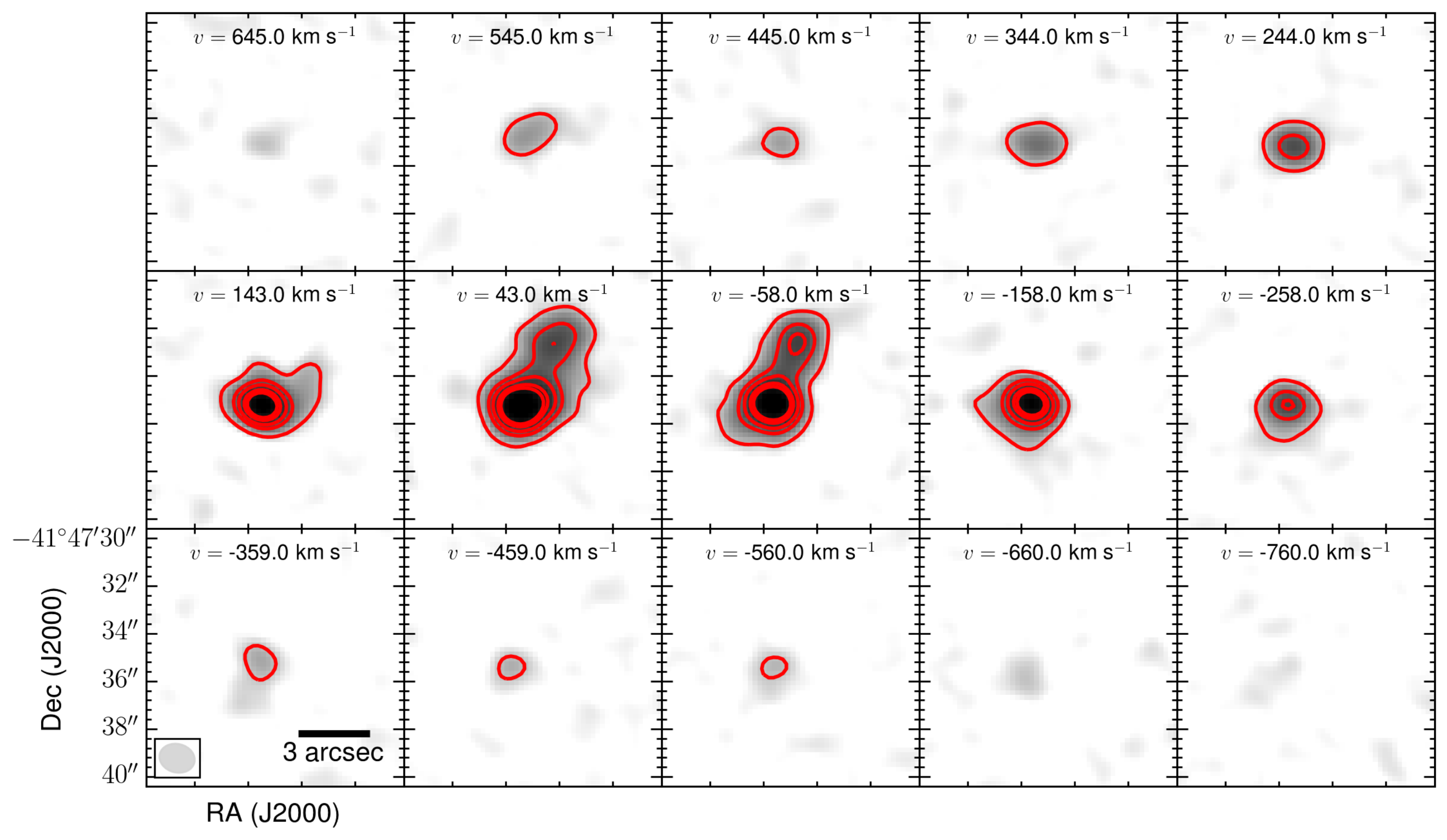}
	\caption{Channel maps for CO($1-0$) emission from IRAS 20100$-$4156, at a velocity resolution of $\sim 100$ \kms. Red contours represent the $\pm 7, 30,50,80,100\sigma$ significance levels. The synthesized beam size is $1''.5\times 1''.2$, and is shown on the bottom left. We spatially resolve the two merging components A and B, in the system, but we do not find a significant velocity offset between them. }
	\label{fig:iras20100_chanmap}
\end{figure*}

\subsubsection{CO($1-0$) emission} \label{sssec:iras20100_co10}

IRAS 20100$-$4156 is a two-component interacting system and we here refer to the two principal components as A and B, respectively (\autoref{fig:iras20100_hst_cont_oh}). The two components have a projected separation of $\sim 2''.5$, corresponding to $5.6$ kpc ($1'' \sim 2.287$ kpc at $z = 0.12975$). We detect and spatially resolve the CO($1-0$) line emission between both components in the continuum-subtracted spectral line cube. The underlying continuum flux is $f_{\rm \lambda} = 1.56 \pm 0.03$ mJy at $\lambda_{\rm obs} = 2.96$ mm, using the line-free channels (\autoref{fig:iras20100_hst_cont_oh}). The dust emission predominantly emerges from component A, though we tentatively detect continuum emission at $\sim 3\sigma$ significance from component B, with a flux of $f_{\lambda} = 93 \pm 36$ $\mu$Jy. We fit the line centroids in spectra extracted from the peak pixels of A and B with 1-D Gaussian line profiles (\autoref{fig:iras20100_allspecs}), and find velocity widths of $\Delta v_{\rm FWHM}^{\rm A} \sim 371.4 \pm 4.2$ \kms$ $ and $\Delta v_{\rm FWHM}^{\rm B} \sim 111.8 \pm 8.2$ \kms, respectively, with both lines centered at $v \sim 0$ \kms. We find no significant velocity/redshift offset between the two components, indicating that they do not show any motion relative to each other. This is also seen in the channel maps for the line core emission (\autoref{fig:iras20100_chanmap}), and in the  moment-1 map (\autoref{fig:iras20100mom10pv}, panel b) which does not show a significant velocity gradient between A and B components. 

The CO moment-0 map (\autoref{fig:iras20100_hst_cont_oh}) is created by integrating over the FWZI of the CO line. We fit the A and B component emission in the CO($1-0$) moment-0 maps with 2D Gaussian components, and find velocity integrated line fluxes of $I^{\rm A}_{\rm CO} = 10.7 \pm 0.4$ Jy \kms$ $ and $I^{\rm B}_{\rm CO} = 1.8 \pm 0.1$ Jy \kms$ $ for components A and B, respectively. These correspond to CO line luminosities of $L'_{\rm CO, A} = (8.4 \pm 0.3) \times 10^{9}$ \lu$ $ and $L'_{\rm CO, B} =(1.4 \pm 0.1) \times 10^{9}$ \lu, respectively. Assuming a CO-H$_{2}$ gas mass conversion factor $\alpha_{\rm CO}= 0.8 $ \acou, suitable for ULIRGs in the local universe \citep[see][for a review]{bolatto2013}, we find total $\rm H_{2}$ gas masses of $ M_{\rm A} = (6.7 \pm 0.2) \times 10^{9} M_{\odot}$ and $ M_{\rm B} = (1.1 \pm 0.1) \times 10^{9} M_{\odot}$ for components A and B respectively. 

We detect outflowing molecular gas from component A in the form of significant emission beyond $v\sim \pm 450$ \kms$ $ (\autoref{fig:iras20100_allspecs}), also seen in the position-velocity (PV) diagram (\autoref{fig:iras20100mom10pv}). We discuss the outflow wing emission in greater detail in the next section. 

\subsubsection{Determination of outflow wings} \label{sssec:iras20100_wings}

The determination of the outflow properties depends sensitively on the velocity widths (ranges) assumed for the high-velocity outflow wings, and we define the velocity ranges for outflow wings in an iterative manner. The initial spectrum is extracted from a $5'' \times 5''$ region centered on the peak pixel, and we simultaneously fit 3 Gaussian profiles to the spectrum, one each for the red wing, the blue wing and the line core, to avoid assumptions about the symmetry of the red and blue wings. The inner limits of the outflow velocity width are defined as the channel at which the wing emission starts dominating over the core emission, and the outer limits are defined using the fitted broad Gaussian components. The cleaned image cube is binned over these velocity ranges to obtain emission maps for the red and blue wings, and CO spectra are re-extracted from the peak emission pixel in the red and blue emission maps. We again fit the obtained spectrum with a sum of 3 Gaussian components, and repeat the same procedure as above until it converges. We bin the calibrated and continuum subtracted $uv$-visibility data over the final velocity ranges, and use model fitting to the binned visibilities for the red and blue wings to determine the outflow positions and fluxes. We fit models to the binned visibility data using the CASA task {\sc uvmodelfit}, assuming a 2-D Gaussian source to obtain the FWHM spatial extent and the flux (see Appendix for details). 

\begin{figure*} 	\centering
	\includegraphics[width=0.66\textwidth]{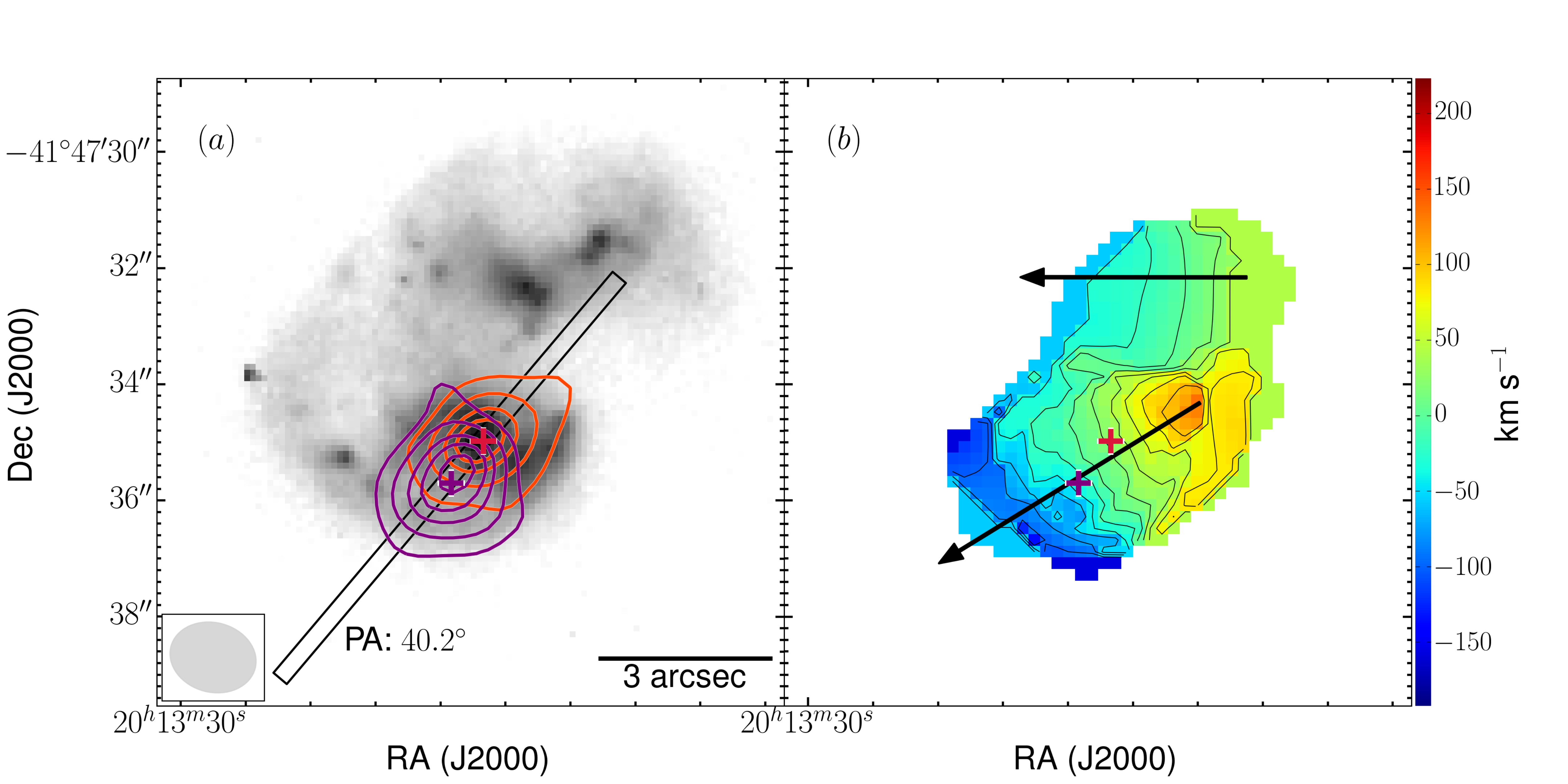}
	\includegraphics[width=0.33\textwidth]{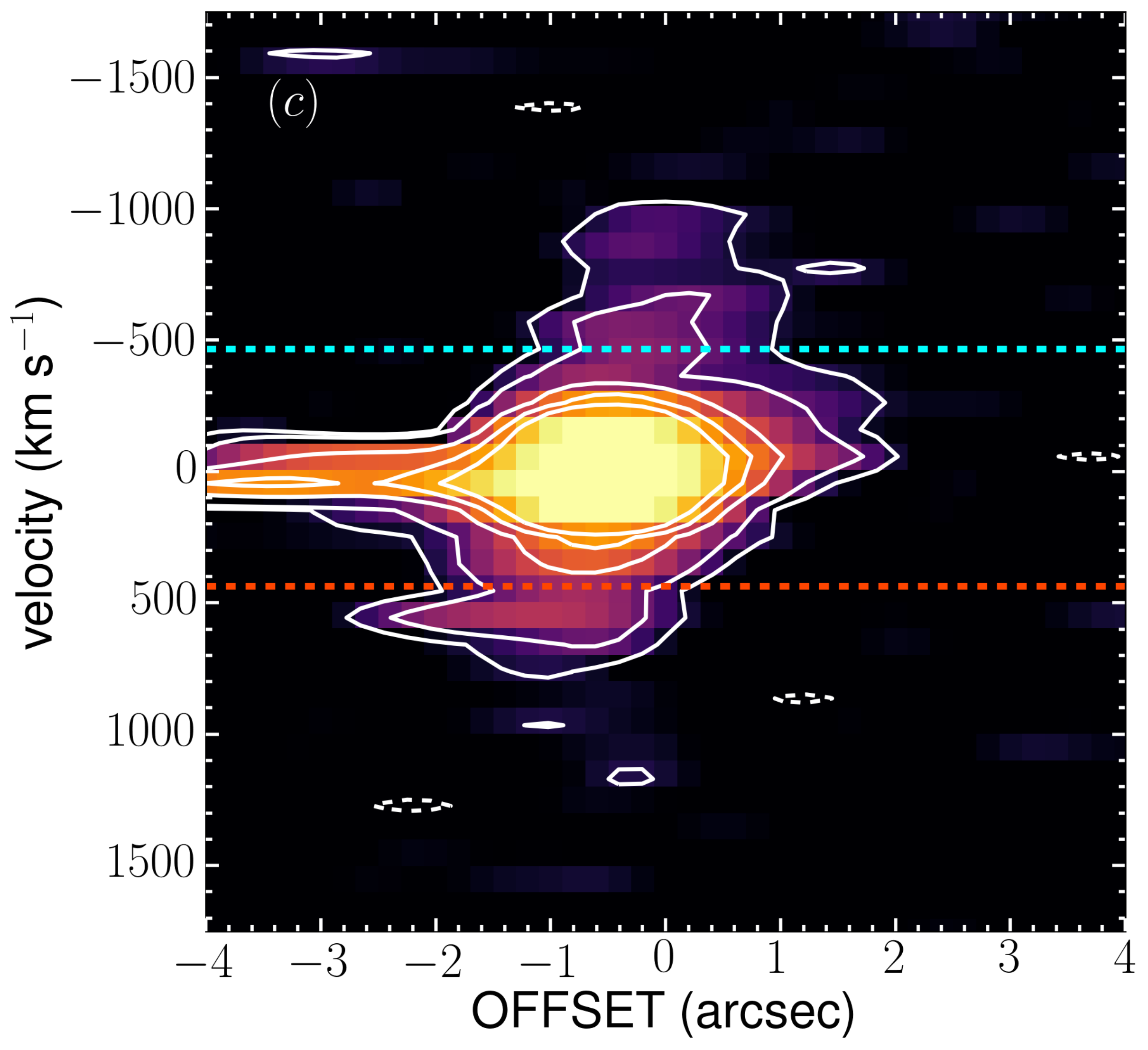}
	\caption{(a) $\pm3,6, 9,11, 13 \sigma$ significance contour levels for the red (shown in red) and blue (shown in purple) outflow emission in IRAS 20100$-$4156, overlaid on the HST/WFC2 814W image. The peaks of the red and blue outflow wing emission (marked by a red and purple cross, respectively) have a projected separation of $0''.9$ ($\sim 2$ kpc) (b) The moment-1 map of the core of the CO($1-0$) line emission ($ -150$ \kms $ < v < 200$ \kms), showing a relatively smooth velocity gradient across both components A and B in IRAS 20100$-$4156 (indicated by black arrows) the contours are marked at increments of $20$ \kms$ $ from $-140$ to $+200$ \kms. The red and blue outflow wings are nearly aligned with the galaxy-wide velocity gradient, though there is no significant velocity gradient between the peaks of the red and blue outflow wing positions (c) Position-velocity (PV) diagram from the region marked in black in (a), along an angle $\sim 40.2 \degree$; the dashed lines show the velocity limits beyond which we detect emission from the red and blue outflow wings. Component B is detected as the extended feature along $v \sim 0$ \kms, with an offset up to $-4''$. The white contours show the $\pm 3, 7, 23, 40, 60\sigma$ significance levels.}
\label{fig:iras20100mom10pv}
\end{figure*}

We find the velocity ranges for the red and blue outflow wings to be $\in (440,1650) $ \kms$ $ and $\in (-1169,-466)$ \kms, respectively. Based on $uv$ model-fitting of the visibility data binned over these velocity ranges, we find integrated fluxes of $I_{\rm CO}^{\rm red} = 0.54 \pm 0.03$ Jy \kms$ $ and $I_{\rm CO}^{\rm blue} = 0.58\pm 0.03$ Jy \kms, and measure a velocity integrated line flux of $I_{\rm CO}^{\rm core} = 11.2 \pm 0.7$ Jy \kms$ $ in the line core. We find a projected spatial offset of $0''.9 \pm 0''.1$ between the peaks of the red and blue emission, which corresponds to a physical distance of $\sim 2.0 \pm 0.2$ kpc and which we define to be the outflow diameter (\autoref{tab:wp}). The peaks of the red and blue wings are spatially offset from the peak CO($1-0$) and dust emission by $0''.3 \pm 0''.05$ and $0''.6 \pm 0.''1$, corresponding to physical distances of $0.7 \pm 0.1$ kpc and $1.4 \pm 0.2$ kpc respectively (\autoref{fig:iras20100mom10pv}). The resultant properties are consistent with those calculated from the cleaned image. The final velocity ranges and fluxes are given in \autoref{tab:wp}. The outflow appears to be aligned with the velocity gradient seen across the galaxy (\autoref{fig:iras20100mom10pv}), although the velocity gradient is small between the peaks of the red and blue outflow wing. This raises the intriguing possibility that rotational support may play a role in the origin of the most powerful outflows \citep[assuming that the outflow is in the plane of the galaxy; see][]{ga2017}.

\subsubsection{CO($3-2$) emission}\label{sssec:iras20100_co32}

We detect CO($3-2$) line emission at a $\sim 21 \sigma$ significance from IRAS 20100$-$4156 (\autoref{fig:iras20100_allspecs}). The A and B components are spatially unresolved in the observations. Using a Gaussian fit to the observed spectrum, we find a linewidth of $\Delta v_{\rm FWHM} = 293 \pm 25$ \kms. Integrating the line emission over the FWZI ($v \sim -269$ \kms$ $ to $v \sim 317$ \kms), we find an integrated flux of $I_{\rm CO (3-2)} = 85.8 \pm 4.0 $ Jy km s$^{-1}$. This corresponds to a line luminosity of $L'_{\rm CO (3-2)}= (7.5 \pm 0.4) \times 10^{9}$ \lu. Using the total observed CO($1-0$) line luminosity (for both the A and B components combined), we find a brightness temperature ratio of $r_{31}\sim 0.8$, consistent with the mean ratio in other local ULIRGs ($ \langle r_{31} \rangle \sim 0.7$, \citealt[][]{papa2012}). 

We do not detect the high-velocity outflow wings seen in CO($1-0$) in CO($3-2$) emission at the depth of these observations. However, we tentatively find that while normalized line profiles for CO($3-2$) and CO($1-0$) agree at $v > 0$ \kms, they show a mismatch at $v < 0$ \kms (\autoref{fig:iras20100_allspecs}, d). This can be quantified using the line luminosity ratio $L'_{\rm CO (3-2)}/L'_{\rm CO(1-0)} = r_{31}$ for the red- and blue-shifted emission in the line core. We find $r_{31}^{\rm blue} = 0.6 \pm 0.2$, while $r_{31}^{\rm red} = 0.9 \pm 0.2$, where we have selected only those channels where CO($3-2$) emission is detected at $\gtrsim 2\sigma$ significance. While both $r_{31}^{\rm red}$ and $r_{31}^{\rm blue}$ are consistent with the median $r_{31}$ in ULIRGs, there is no physical reason for a difference in excitation between the red and blue-shifted emission. The observations can be explained if CO($3-2$) emission is under-luminous on the blue-shifted side as compared to the redshifted emission. As the CO($3-2$) emission traces more compact nuclear emission than CO($1-0$), self-absorption in the molecular gas can result in decreased flux, leading to a lower $r_{31}^{\rm blue}$ than $r_{31}^{\rm red}$. We therefore conclude that the average $r_{31} \gtrsim 0.8$, and treat it as a lower limit on the gas excitation. 

\subsubsection{OHM emission}\label{sssec:iras20100_ohm}

We detect strong OHM emission from IRAS 20100$-$4156 at a $\sim 40 \sigma$ significance. We detect line emission from a velocity range $-927$ \kms$ $ to 527 \kms, with tentative evidence of absorption redward of 527 \kms (\autoref{fig:iras20100_allspecs}). The OHM emission could be contaminated by the OH 1.665 GHz satellite line (OH $^{2} \Pi_{3/2}$, $ J=3/2, F=1^e-1^f$) at the expected position of $v\sim + 360$ \kms, but we do not detect excess emission at that velocity. As the emission is spatially unresolved, we extract the spectrum from the peak pixel of the moment-0 map (\autoref{fig:iras20100_allspecs}). Based on a 2-D elliptical Gaussian fit to the moment-0 map, we find an integrated line flux of $I_{\rm OH} = 27.9 \pm 0.8$ Jy km s$^{-1}$. This corresponds to a line luminosity of $L_{\rm OH} = (1.5 \pm 0.1) \times 10^{4} L_{\odot}$. We find a systematic velocity offset of $\delta v \sim 190 \pm 7$ \kms$ $ between the CO($1-0$) and OHM lines, as well as a spatial offset of $0''.3 \pm 0''.1 $ between the peaks of emission. We discuss these differences further in \autoref{ssec:ohm}. 

\subsection{IRAS 03158$+$4227} \label{ssec:iras03158}

\subsubsection{CO($1-0$ emission)} \label{sssec:iras03158_co10}

\begin{figure*}
	\includegraphics[width=\textwidth]{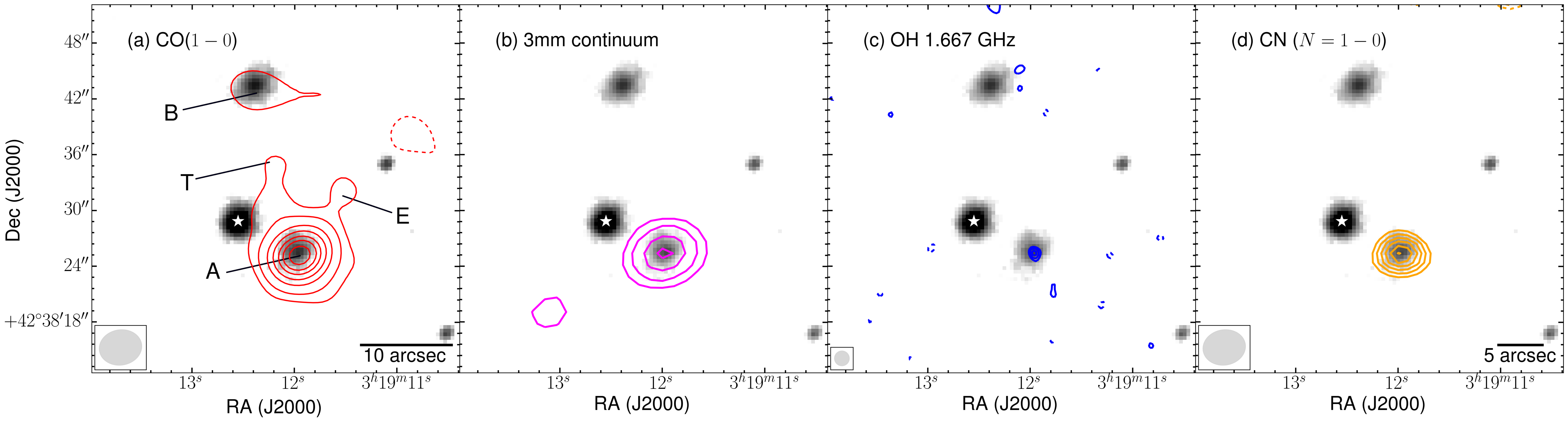}
	\caption{(a) Moment-0 map of CO($1-0$) emission from IRAS 03158$+$4227, created by integrating over the velocity range ($-944, 1750$) \kms, overlaid on the SDSS $g'$ band image (b) the 3mm continuum emission from IRAS 03158$+$4227 (c) Moment-0 map of the spatially unresolved OHM emission, created using the velocity range ($-1404, 296$) \kms.(d) Moment-0 map of the CN($N = 1-0$) emission, created using the velocity range ($-289,429$) \kms. Panels (a) and (b) show contours marked at $\pm 3,7,15,23,32,40,50,60,80\sigma$ significance levels, while (c) and (d) show the contours at $\pm 3,4,5,6,7,8 \sigma$ significance levels. The peaks of the two components of the interacting pair (marked A and B) have a projected separation of $18''.9$ (45.4 kpc). We detect the molecular gas outflow from component A. Synthesized beam sizes are $4''.8\times 3''.9$ for (a), (b), and (d) and $1''.6 \times 1''.3$ for (c), and are shown at the bottom left of each panel.} The white star shows the foreground star in the field of view.
\label{fig:iras03158_allimgs}
\end{figure*}

\begin{figure}
	\centering
	\includegraphics[width=0.45\textwidth]{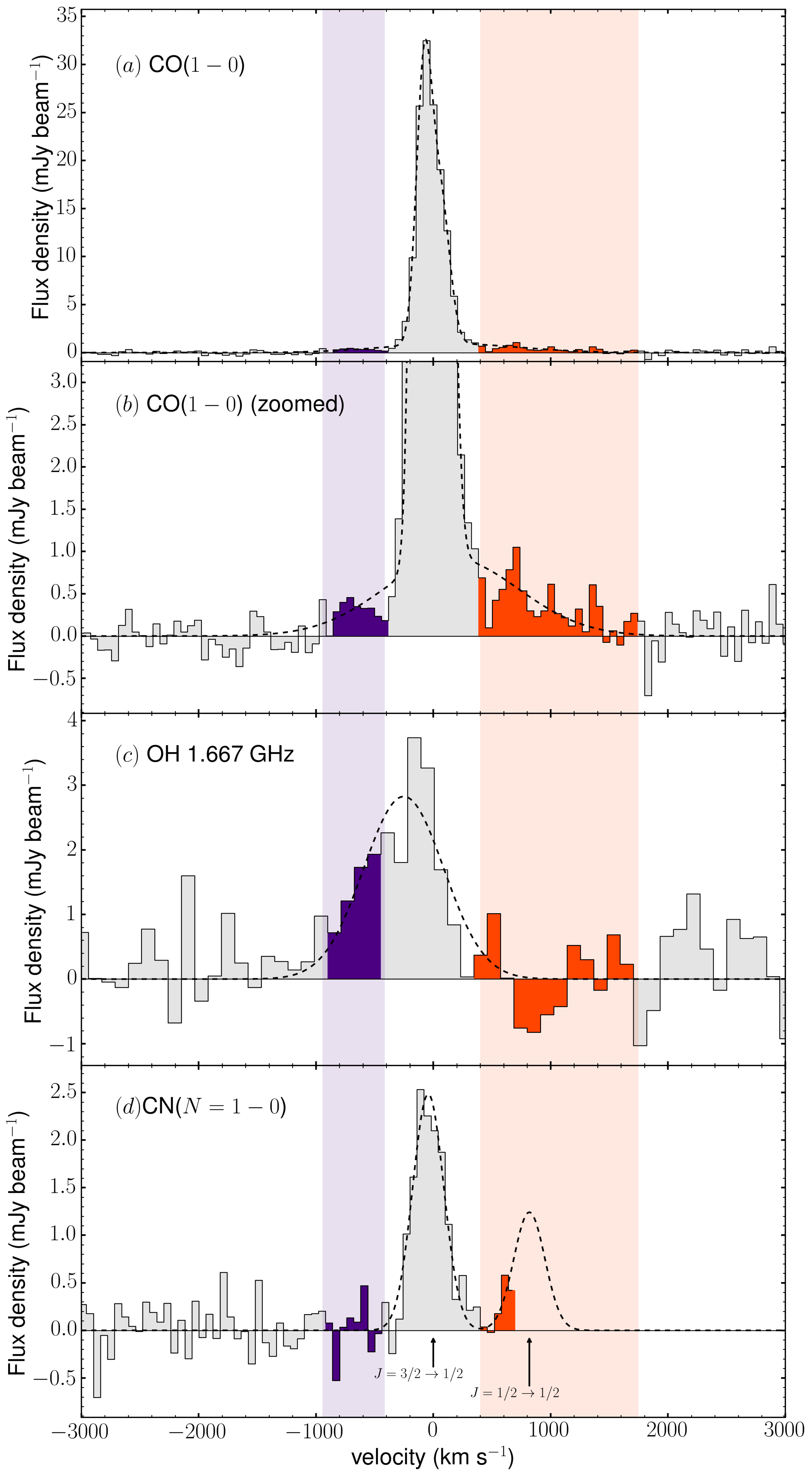}
	\caption{(a) CO($1-0$) spectrum extracted from the peak pixel of component A, IRAS 03158+4227 (b) Same as (a), zoomed in to show the high-velocity wings (c) OHM spectrum for IRAS 03158+4227, extracted from the peak emission position (d) CN($N=1-0$) spectrum for IRAS 03158+4227, with only the $J=3/2-1/2$ component fully detected. The dashed line show the best fit spectra using a sum of three 1D Gaussian components for CO($1-0$) and using a one-component Gaussian fit for OHM emission. For CN($N=1-0$), the dashed line shows the expected line profile for the doublet, assuming LTE. The purple and orange regions show the velocity ranges for the high-velocity outflow wings in CO($1-0$), defined in \autoref{sssec:iras20100_wings}, and $v = 0$ \kms$ $ has been defined using $z_{\rm CO}$. }
\label{fig:iras03158_allspecs}
\end{figure}

\begin{figure*}
	\centering
	\includegraphics[width=\textwidth]{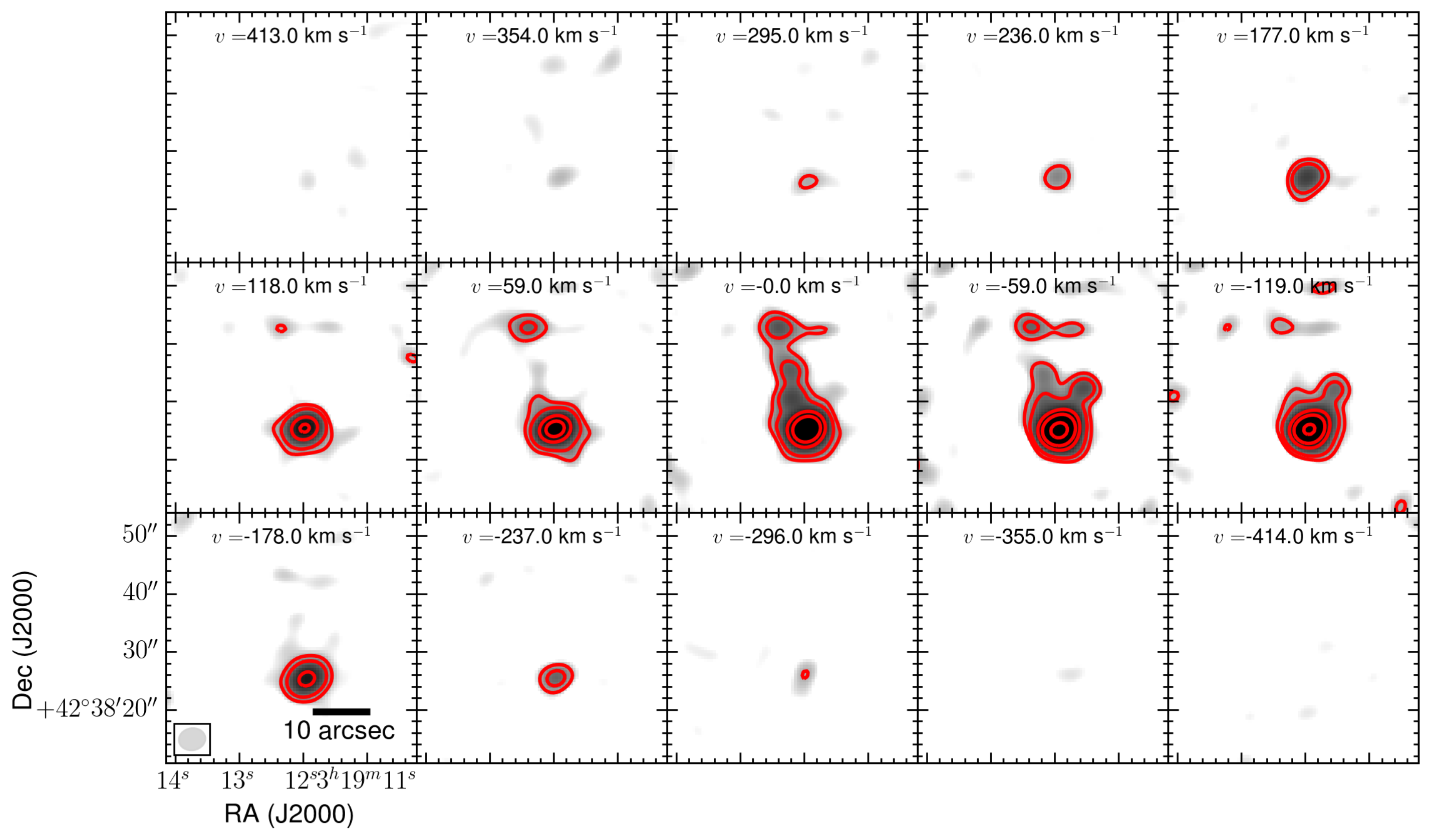}
	\caption{Channel maps for CO($1-0$) emission from IRAS 03158+4227, with a velocity resolution of $\sim 60$ \kms. The red contours represent the $\pm 7, 30,50,80,100\sigma$ significance levels. The synthesized beam size is $4''.8\times 3''.9$, and is shown on the bottom left. We spatially resolve multiple components of the merging system - A, B, T and E, and note that there is no significant velocity offset between these components.}
	\label{fig:iras03158_chanmap}
\end{figure*}

\begin{figure*} 	\centering
	\includegraphics[width=0.66\textwidth]{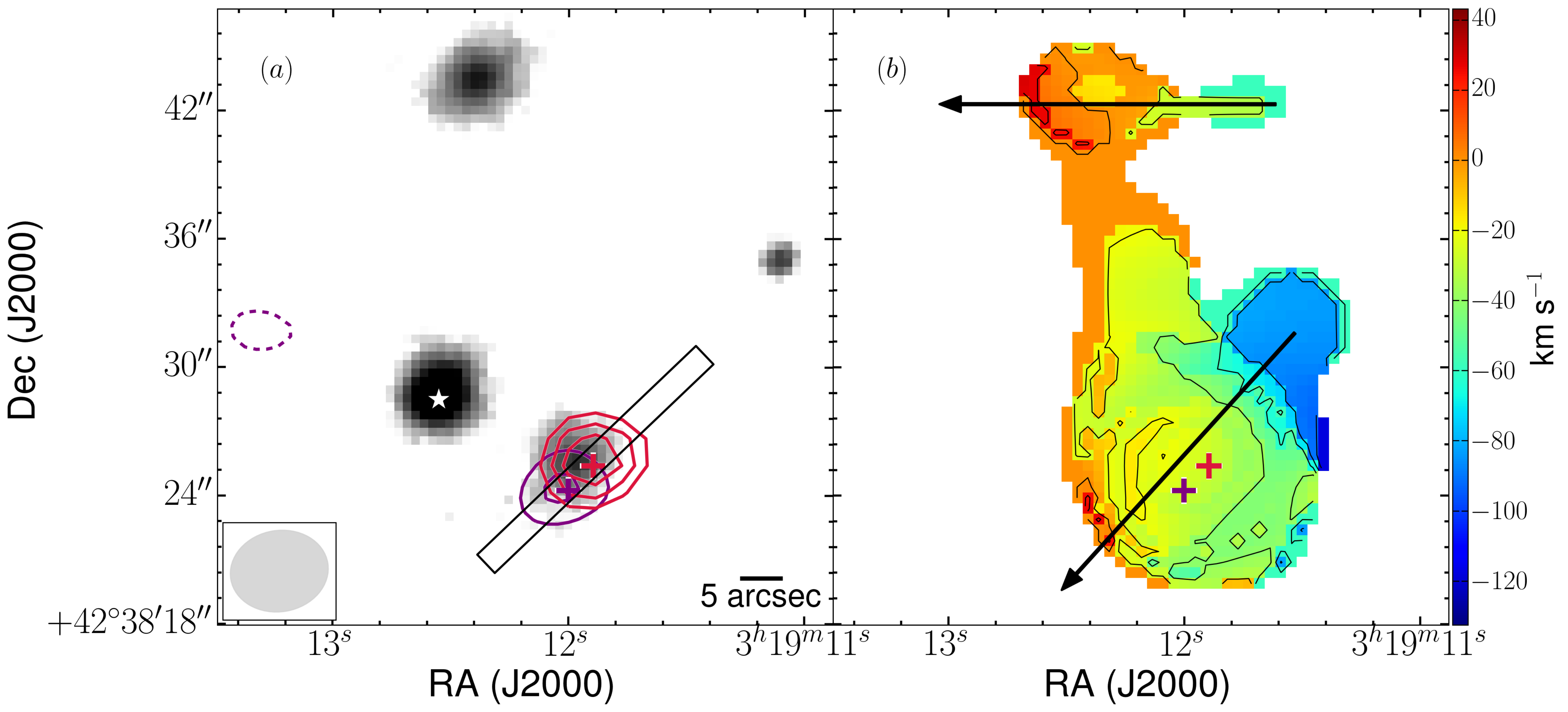}
	\includegraphics[width=0.33\textwidth]{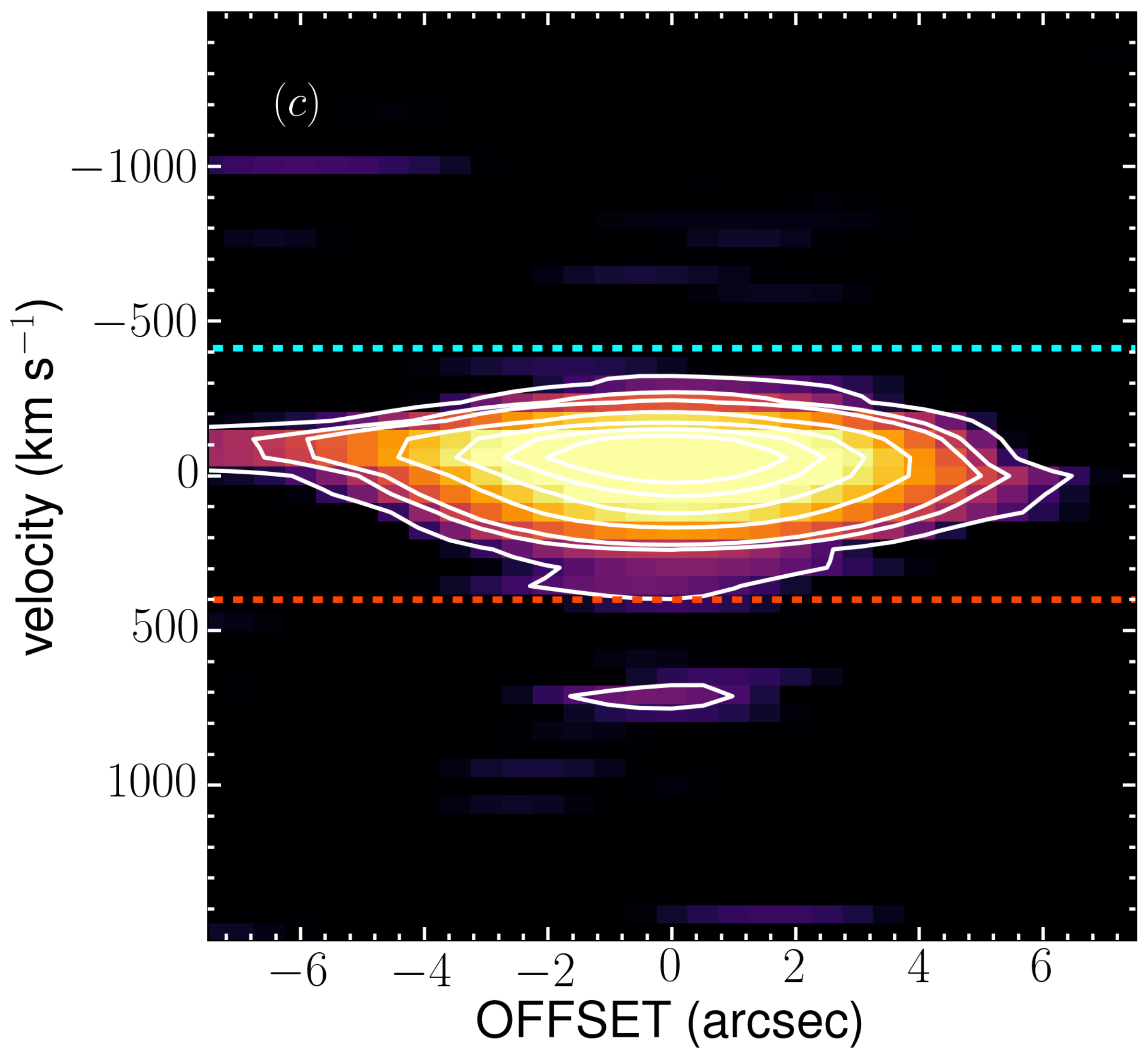}
	\caption{(a) $\pm 3, 6, 9,11, 13 \sigma$ significance contour levels for the red (shown in red) and blue (shown in purple) outflow emission in IRAS 03158+4227, overlaid on the SDSS $g'$ band image. The peaks of the red and blue outflow wing emission (marked by a red and purple cross, respectively) have a projected separation of $1''.6$ ($\sim 3.9$ kpc) b) The moment-1 map of the core of the CO($1-0$) line emission ($ -120$ \kms $ < v < 40$ \kms), which shows distinct velocity gradients along both components A and B, as shown by the arrows in each; the contours are marked at increments of $20$ \kms from $-120$ to $+120$ \kms (c) The PV diagram extracted from the region marked in (a), along an angle $\sim 46 \degree$; the dashed lines show the velocity ranges for the red and blue outflow wings, as defined in \autoref{sssec:iras20100_wings}. The peaks of the red and blue outflow wing emission are nearly anti-aligned with the large scale rotation of the galaxy (indicated by black arrows for both component A and B), though there is no significant velocity gradient between the peaks of the red and blue outflow wing positions. Component E is detected as the extended feature detected along $v \sim 0$ \kms. The white contours show the $\pm 3,7,10,23,40,60\sigma$ significance levels.}
		\label{fig:iras03158mom10pv}
\end{figure*}

IRAS 03158$+$4227 is a merging system with two distinct galaxies spatially offset by $18''.9$, corresponding to $45.4$ kpc ($1'' \sim 2.360$ kpc at $z = 0.13459$) as well as tidal structure with two components, one connecting the two galaxies. The two principal components are labeled A, B, while the two tidal components are labeled T and E (\autoref{fig:iras03158_allimgs}). We detect and spatially resolve the CO($1-0$) line emission between each of these, and detect high-velocity outflowing molecular gas from component A (\autoref{fig:iras03158_allspecs}). We do not find a significant velocity offset between any of the components, as can be seen in the channels maps of the CO emission (\autoref{fig:iras03158_chanmap}), and the moment-1 map of the emission shows a relatively smooth velocity gradient across all components (\autoref{fig:iras03158mom10pv}). The moment-0 map of the CO emission is made by integrating over the FWZI line width (\autoref{fig:iras03158_allimgs}). The continuum emission, restricted to component A, has a flux of $f_{\lambda} = 0.94 \pm 0.03 $ mJy at $\lambda_{\rm obs} = 3$ mm, using the line-free channels. 

The velocity widths and spatial extents of components A, B, T and E are significantly different. We therefore extract the fluxes for each component in the following iterative manner; we select a $3''\times 3''$ region around the A, B and E components (and a $9''\times 3''$ region around the elongated tidal tail T). We extract spectra from each of these regions, which is well described by a Gaussian line profile. Based on these spectra, we integrate over all contiguously positive channels around the line center to obtain a moment-0 map for each region. We then fit each component with a 2-D elliptical Gaussian, and thus extract the flux from each component - A, B, T and E. We iteratively fit and subtract emission from A, B and T, and find velocity integrated line fluxes of $I_{\rm CO}^{\rm A} = 13.1 \pm 0.1$, $I_{\rm CO}^{\rm B} = 2.8 \pm 0.3$, $I_{\rm CO}^{\rm T} = 2.1 \pm 0.4$ and $I_{\rm CO}^{\rm E} = 1.2 \pm 0.1$ Jy \kms$ $ respectively for each of the components. These imply line luminosities of $L'_{\rm CO, A} = (1.1\pm0.1) \times 10^{10}$, $L'_{\rm CO, B} = (2.4 \pm0.2)\times 10^{9}$, $L'_{\rm CO, T} = (1.8 \pm 0.4) \times 10^{9}$, and $L'_{\rm CO, E} = (9.7 \pm 0.8) \times 10^{8}$ \lu$ $ respectively. Assuming a CO-luminosity to $\rm H_{2}$ gas mass conversion factor of $\alpha_{\rm CO} = 0.8$ \acou$ $ for A and B, and a $\alpha_{\rm CO} = 3.6$ \acou$ $ for the tidal components, we find gas masses of $M_{\rm A} = (8.9 \pm 0.8 ) \times 10^{9}$, $M_{\rm B} = (1.9 \pm 0.2) \times 10^{9}$, $M_{\rm T} = (6.3 \pm 1.3) \times 10^{9}$ and $M_{\rm E} = (3.6 \pm 0.5) \times 10^{9} M_{\odot}$, respectively. The differing conversion factors are physically motivated, with the lower conversion factor suitable for ULIRGs, which typically show higher gas temperatures, while the higher \aco$ $ is more realistic for molecular gas in MW-type giant molecular clouds, and therefore is more suitable for the tidal tails. We find a gas-mass ratio of $\sim 5:1 $ between the two principal components A and B. We detect the high-velocity wing emission at velocities beyond $\pm 1000$ \kms$ $ in the PV diagram for IRAS 03158+4227 (\autoref{fig:iras03158mom10pv}). 

\subsubsection{Determination of outflow wings} \label{sssec:iras03158_wings}

We obtain the velocity widths for the outflow wings in the same manner as for IRAS 20100$-$4156 (see \autoref{sssec:iras20100_wings}), and find the velocity range $\in (400,1750) $ \kms$ $ as the red and $\in (-944,-413)$ \kms$ $ as the blue outflow wings respectively, for IRAS 03158+4227. Based on $uv$ model-fitting of the binned visibility data over this velocity range, we find integrated fluxes of $I_{\rm CO}^{\rm red} = 0.80 \pm 0.22$ Jy \kms$ $ and $I_{\rm CO}^{\rm blue} = 0.31\pm 0.15$ Jy \kms$ $ in the red and blue wings respectively. We measure a total velocity integrated line flux of $I_{\rm CO}^{\rm core} = 8.4 \pm 0.7$ Jy \kms$ $ in the line core. We find a spatial offset of $1''.6 \pm 1''.1$ between the peaks of the red and blue emission, which corresponds to a physical distance of $3.9 \pm 2.7$ kpc (which we use as the outflow diameter). The peaks of the red and blue wings are spatially offset from the CO($1-0$) and dust peaks by $0''.6 \pm 0''.2$ and $1''.1 \pm 0''.5$ respectively, corresponding to a physical distance of $\sim 1.4 \pm 0.4$ kpc and $\sim 2.6 \pm 1.2$ kpc (\autoref{fig:iras03158mom10pv}). The difference in the red and blue outflow peak offsets indicates a significant asymmetry in the outflow morphology. The outflow is counter-aligned with the galaxy-wide velocity gradient (\autoref{fig:iras03158mom10pv}), in direct contrast with that seen in IRAS 20100$-$4156. It is therefore unlikely that host galaxy rotation is an essential component in launching the outflows.

\subsubsection{CN line emission} \label{sssec:iras03158_cn}

CN abundance is enhanced in UV-irradiated surfaces of molecular gas clouds, and therefore preferentially traces photo-dissociated regions (PDRs; \citealt{rf1998}), while CO traces the entirety of the molecular gas (dense as well as diffuse components). A low $L'_{\rm CO}/L'_{\rm CN}$ ratio therefore indicates relatively more PDR/XDR dominated molecular gas. 

We detect CN($N=1-0$) line emission from only component A in IRAS 03158+4227 \autoref{fig:iras03158_allimgs}. While the line is a doublet, with the fine structure (FS) components ($J = 1/2 - 1/2$) and ($J = 3/2 - 1/2$) separated by $\delta \nu_{\rm rest} \sim 0.31$ GHz, corresponding to $\delta v \sim 819$ \kms, we cover only the ($J = 3/2 - 1/2$) component of the transition. Fitting a 1D Gaussian to the spectral line, we find a velocity width of $\Delta v_{\rm FWHM} = 312.5 \pm 24.9$ \kms, centered on $v_{\rm sys} = -42.9 \pm 10.4$ \kms. We create a moment-0 map by integrating over the FWZI velocity range ($v = -289$ \kms$ $ to $v = 429$ \kms). Fitting the moment-0 emission with a 2-D Gaussian source, we find an integrated line flux of $I_{\rm CN} = 1.0 \pm 0.1$ Jy \kms. This corresponds to a line luminosity of $L'_{\rm CN} = (8.9 \pm 0.5) \times 10^{8}$ \lu. 
 
Under conditions of Local Thermal Equilibrium (LTE) between the two FS components, we expect that the two components will have a ratio of

\begin{equation} \label{eq:1}
R_{12} \mathrm{CN}(N = 1 - 0) = \frac{I(J = 1/2 - 1/2)}{I(J = 3/2 -1/2)} = 0.5 
\end{equation}

The expected line profile of the doublet in LTE is shown in \autoref{fig:iras03158_allspecs}. While the CN($N=1-0, J=3/2-1/2$) and ($N=1-0,J=1/2-1/2$) lines are split further into hyperfine structure lines (HFS), we do not spectrally resolve them. Assuming LTE, we find an expected integrated line flux of $I_{\rm CN} = 1.5 \pm 0.2$ Jy \kms$ $ for both components together, corresponding to a line luminosity of $L'_{\rm CN} = (1.3 \pm 0.1) \times 10^{9}$ \lu. Using the CO($1-0$) line luminosity as calculated above, we find a line luminosity ratio of $L'_{\rm CO}/L'_{\rm CN} \sim 8.2 \pm 0.5$. The observed $L'_{\rm CO}/L'_{\rm CN}$ is consistent with the globally averaged line-ratios seen in other ULIRGs and is similar to that seen in Mrk 231 ($L'_{\rm CO}/L'_{\rm CN} \sim 8$, \citealt{aalto2002, peres2007}). With a critical density $5$ times lower than HCN($1-0$), CN($N=1-0$) does not trace the same molecular gas as HCN, and $L'_{\rm HCN}/L'_{\rm CN}$ varies in the range $\sim 0.5-6$ for galaxies with a constant $L'_{\rm CO}/L'_{\rm HCN}$ \citep{aalto2002}. We therefore do not use it to calculate the dense gas mass. 

\subsubsection{OHM emission} \label{sssec:iras03158_ohm}

We detect spatially unresolved OHM emission from component A of IRAS 03158$+$4227 (\autoref{fig:iras03158_allspecs}). Fitting a 1D Gaussian to the spectral line profile, we find a velocity width of $\Delta v_{\rm FWHM} \sim 803 \pm 110$ \kms, and a velocity offset of $v_{0} \sim -252 \pm 46$ \kms$ $ between the CO and OHM lines. Based on a 2-D elliptical Gaussian fit to the moment-0 map (\autoref{fig:iras03158_allimgs}), we find an integrated line flux of $I_{\rm OH}= 4.1 \pm 0.8$ Jy \kms. This corresponds to a line luminosity of $L_{\rm OH} = (2.4 \pm 0.5) \times 10^{3} L_{\odot}$. We detect a marked asymmetry in the maser emission line profile, with a significant excess in the blueshifted emission, similar to that observed in the OHM spectrum for IRAS 20100$-$4156. This is discussed further in \autoref{ssec:ohm}. 

\section{Analysis} \label{sec:analysis}
\subsection{Outflow properties}\label{ssec:outflows}

The physical properties of outflows can be encapsulated by the mass outflow rate $\dot M_{\rm OF}$, the momentum outflow flux $\dot P_{\rm OF}$, and the kinetic flux $\dot E_{\rm kin,OF}$ of the outflowing gas, and the use of all three is important to distinguish between different outflow models \footnote{All outflow properties are identified with a subscript $\rm OF$}. These are defined as 

\begin{align} \label{eq2}
\dot M_{\rm OF} &= v_{\rm OF} \Omega R_{\rm OF}^{2} \rho_{\rm OF} \\
\dot P_{\rm OF} &= v_{\rm OF} \dot M_{\rm OF} \\
\dot E_{\rm kin, OF} &= \dot M_{\rm OF} v_{\rm OF}^{2} /2 
\end{align}

where $\dot M_{\rm OF}$ is the mass outflow rate, $v_{\rm OF}$ is the outflow velocity, $R_{\rm OF}$ is the outflow radius, $\Omega$ is the solid angle of the outflow ($4\pi$ steradians for a spherical outflow) and $\rho_{\rm OF}$ is the density of the outflowing material at $R_{\rm OF}$. In the case of a spherical outflow of radius $R_{\rm OF}$ uniformly filled with outflowing clouds, assuming that $\rho_{\rm OF} \sim \rho_{\rm avg}/3$ (where $ \rho_{\rm avg}$ is the average density of the outflowing gas), and that the mass in the outflow is $M_{\rm OF} = (\Omega /3) R_{\rm OF}^{3} \rho_{\rm avg}$, the mass outflow rate \mout$ $ in steady state is given by 

\begin{equation} \label{eq:3}
\dot M_{\rm OF} = M_{\rm OF} \frac{v_{\rm OF}}{R_{\rm OF}}
\end{equation}

However, for a thin shell-like geometry of the outflow, the volume density is given by 

\begin{equation} \label{eq:4}
\rho_{OF} = \frac{3M_{\rm OF}}{\Omega (R_{\rm in}^{3} - R_{\rm OF}^{3})} \approx \frac{M_{\rm OF}}{\Omega R_{\rm OF}^{2} \Delta R} 
\end{equation}

where $R_{\rm in}$ and $R_{\rm OF}$ are the inner and outer radii of the outflowing shell, and $\Delta R$ is the thickness of the shell. In this geometry, the mass outflow rate \mout$ $ is given by 

\begin{equation} \label{eq:5}
\dot M_{\rm OF} = M_{\rm OF} \frac{v_{\rm OF}}{\Delta R}
\end{equation}

Note that the outflow solid angle is included in $M_{\rm OF}$, and thus \autoref{eq:5} holds for spherical as well as biconical outflows. The thin shell case implies higher outflow rates, as $\Delta R$ is smaller than $R_{\rm OF}$. This can also be thought of as the instantaneous estimate of the energetics, valid over timescales $\Delta \tau \sim \Delta R/v_{\rm OF}$; the time-average of this expression yields the same expression as the steady-state case described above; \citealt{ga2017}). In order to distinguish between these two cases observations with a higher spatial resolution (at least $\sim 100$ pc) are required. We therefore calculate the outflow properties assuming the first model (\autoref{eq:3}). This is consistent with the formalism adopted in \citet{heckman2015, thompson2015, ga2017}, but is a factor of $3$ lower than the values assumed in \citet{feruglio2010, maiolino2012, cicone2014, garcia2015}. The decreasing density as a function of radius assumed in the model used is also consistent with observations that outflowing dense gas is more spatially compact than ground-state CO emission \citep{feruglio2015}.

We obtain outflowing gas masses using \aco$ $ to convert between the line luminosity and the gas mass, assuming \aco $\sim 0.8$ \acou$ $ to calculate the outflow mass (a typical value for ULIRGs), and assuming that the outflowing gas is 100\% molecular. The value of \aco$ $ can be lower by a factor of $\sim 2.4$ or higher by a factor of $\sim 5.5$, depending on the two limiting cases of optically thin molecular gas \aco $ = 0.34$ \acou, vs the upper limit of a MW-like \aco $= 4.3$ \acou \citep[see][for a review]{bolatto2013}. In fast moving winds/outflows, CO emission is expected to be optically thin i.e. \aco $\gtrsim 2 \times$ lower than our assumed value, which has been seen in radio-jet driven molecular outflows \citep{osterloo2017}. \aco $ $ could also be low if the outflowing molecular gas (which is mass-loaded early in the outflow; \citealt{ga2017}) is enriched by the nuclear starburst and/or by \emph{in situ} star formation in the outflow \citep{maiolino2017}. However, observations of the CO excitation in Mrk 231 find the excitation in the outflowing gas to be similar to that seen in the bulk of the molecular gas, arguing for a similar \aco$ $ in the outflow and the rest of the galaxy \citep{cicone2012}. Theoretical modeling of the multi-phase ISM in outflows by \citet{richings2017a,richings2017b} suggests that the outflowing gas could have \aco$ $ as much as $5.3$ times lower than our assumed value. However, they also assume a molecular gas fraction of $\sim 20\%$ instead of 100\%, and the two assumptions together cancel out to give an \aco$ $ close to our adopted value (within $\sim 10\%$). However, we emphasize that there is a systematic uncertainty in the outflow mass, momentum flux, and energy flux due to \aco, and further observations of isotopologues such as $^{13}$CO are urgently needed to determine its appropriate value.

We use the mass-weighted outflow velocity, averaged between the red and blue outflow wings, as the representative outflow velocity $v_{\rm OF}$. We do not correct for inclination effects so as to get the most conservative outflow properties; $\dot M_{\rm OF}$ is independent of inclination, while both $\dot P_{\rm OF}$ and $\dot E_{\rm kin, OF}$ are lower limits. We assume the radius of the outflow ($R_{\rm OF}$) to be half the separation of the peaks of the red and blue emission peak, as described in \autoref{sssec:iras20100_wings}, \autoref{sssec:iras03158_wings}. We define the molecular gas depletion timescale to be $\tau_{\rm OF}^{\rm dep} = M_{\rm gas}/ \dot M_{\rm OF}$, assuming that the outflow rate remains constant. The final outflow properties are listed in \autoref{tab:TableOutflow}. 

\subsection{Determining the AGN luminosity}\label{ssec:agnfrac}

Both outflow sources are highly IR luminous ($L_{\rm IR}> 10^{12} L_{\odot}$), with the bulk of the IR luminosity emerging from their compact, dust-obscured nuclei. Identifying what fraction of this IR luminosity is due to the AGN or the compact nuclear starburst is essential for testing observed outflow parameters against theoretical models for the same. We therefore use Spectral Energy Distribution (SED) fitting, mid-IR diagnostics, and archival X-ray observations to determine $L_{\rm AGN}$. The caveats to all these are discussed in \autoref{ssec:caveats}. 

\subsubsection{SED fitting} \label{sssec:sedfitting}

We perform Spectral Energy Distribution (SED) fitting for both sources using CIGALE (Code Investigating GAlaxy Emission; \citealt{noll2009, serra2011}). CIGALE builds galaxy SEDs from UV to radio wavelengths assuming a combination of modules and summing the emission from each, and allows modeling the star formation history (SFH), the stellar emission using population synthesis models \citep{bruzual2003, maraston2005}, nebular lines, dust attenuation \citep[e.g.,][]{calzetti2000}, dust emission \citep[e.g.,][]{draine2007, casey2012}, contribution from AGN \citep{dale2014,fritz2006}, and radio synchrotron emission. The SEDs implicitly maintain the energy budget, with consistency between UV dust attenuation and the dust emission. For our SED fitting, we assume the star formation history to be a double exponentially decreasing function of time, the dust attenuation as in \cite{calzetti2000}, and the dust emission models from \cite{dl2014}. 

We use all available photometric data points (\autoref{tab:phot}) for each of the sources\footnote{We exclude the IRAC 3.6 - 8 \mum photometry for IRAS 03158+4227 as the IRS map shows a mispointing.} as described in Section \autoref{ssec:arch}. We add multiple line-free continuum points from the IRS spectra/PACS spectroscopy to our SED fitting, to ensure consistency between the spectroscopic and the SED fit observations. 

\begin{figure}
\centering
\includegraphics[width=0.5\textwidth]{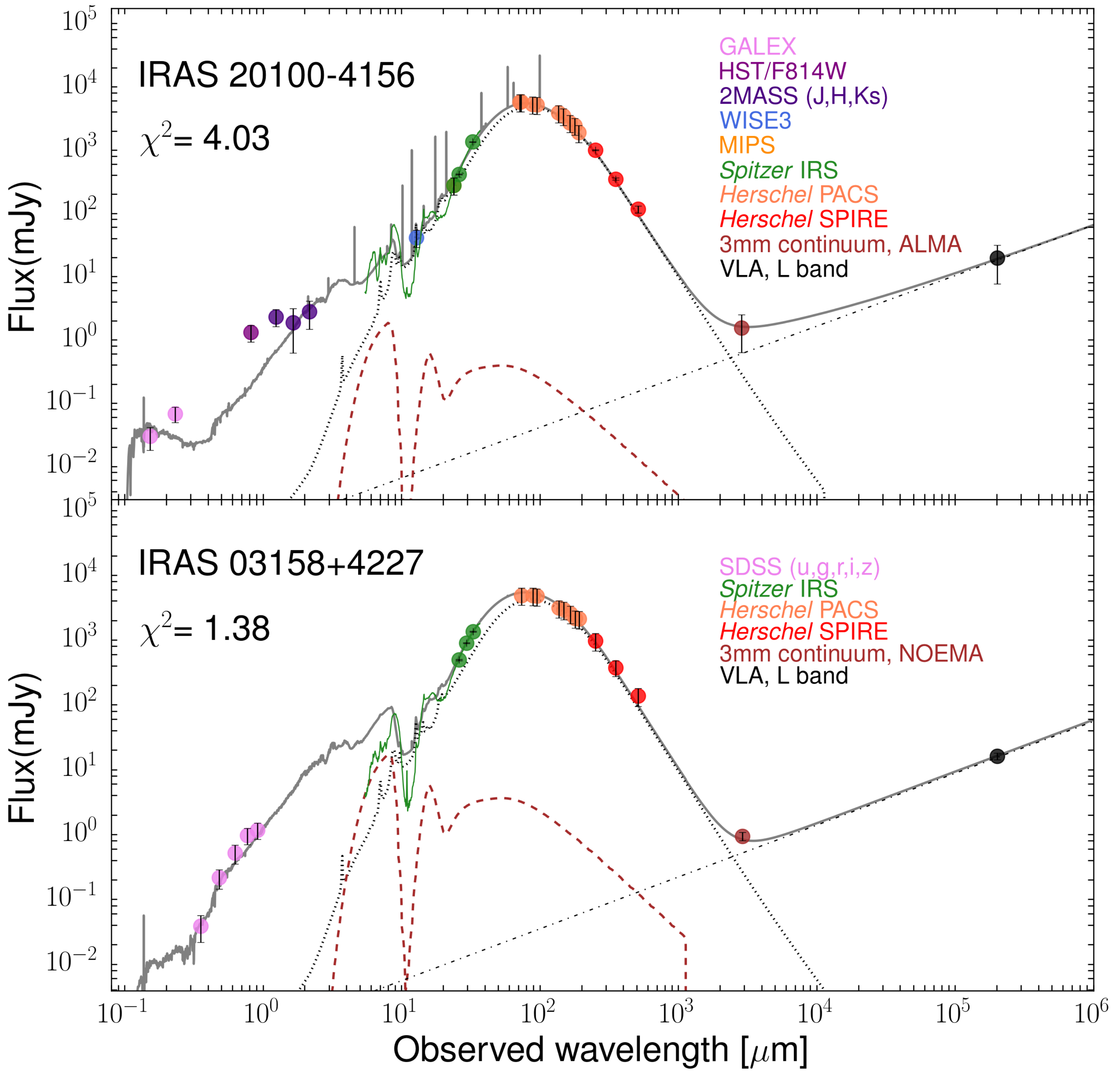}
\caption{Best-fit Spectral Energy Distributions (SEDs) for IRAS 20100$-$4156 and IRAS 03158$+$4227 obtained using CIGALE. The solid grey line shows the overall best fit, the dotted line shows the dust emission SED, the dashed brown line shows the AGN emission, the dash-dotted line shows the radio continuum, and the green line shows the observed \emph{Spitzer} IRS spectra.}
\label{fig:iras20100_cigale}
\end{figure}

The photometry resolves the targets and their merging/interacting companions in some cases, and not in others. For IRAS 20100$-$4156, the A and B components are not dynamically independent and are only spatially resolved in the HST images. We therefore treat them as one system. We re-extract HST photometry so as to include the contribution from both A and B components, and then perform the SED fitting. Based on the modest detection of the $3$ mm dust continuum from the component B, we expect that the bulk of the FIR emission is from component A. For IRAS 03158$+$4227, the A and B components are separated by a projected distance of $\sim 22''$ and spatially resolved in all images except \emph{Herschel} SPIRE photometry. We therefore perform the SED fitting for A with and without the SPIRE photometry, to test for contamination from component B, and find no significant differences in our results. The near-IR wavelengths for IRAS 03158+4227 ($< 6$ $\mu$m) also show contamination from its neighboring star, and we re-extract the photometry where possible. Finally, CIGALE performs a probability distribution function analysis for our specified module parameters and obtains the likelihood-weighted mean value for each. The best-fit SED models are shown in \autoref{fig:iras20100_cigale}. 

We test for any AGN contribution to the SEDs using the \citet{fritz2006} module, which includes AGN emission reprocessed by a dusty torus, with the torus opening angle, orientation, inner and outer radii being free parameters (see \citealt{fritz2006} for further details). We extensively test for AGN contributions to the IR luminosity by varying the initial conditions over the entire allowed parameter space for dusty torus models. We find bolometric AGN luminosities of $L_{\rm AGN} = (7.2 \pm 0.3) \times 10^{11} L_{\odot}$ and $L_{\rm AGN} = (1.3 \pm 0.1) \times 10^{12} L_{\odot}$ for IRAS 20100$-$4156 and IRAS 03158+4227, and $L_{\rm IR}=3.7 \times 10^{12} L_{\odot}$ and $L_{\rm IR} = 3.8 \times 10^{12} L_{\odot}$. Together these correspond to AGN fractions of $f_{\rm AGN} = 0.2 \pm 0.1 $ and $f_{\rm AGN} = 0.3 \pm 0.1$ for IRAS 20100$-$4156 and IRAS 03158+4227, respectively, indicating that neither of the sources is AGN-dominated. The final parameters estimated from the SED fitting - SFR, dust mass, and stellar mass - are listed in \autoref{tab:sedfits}.

We find a gas to dust mass ratio $\delta_{\rm GDR} \sim 10-20$ for both IRAS 20100$-$4156 and IRAS 03158+4227. Previous values for $\delta_{\rm GDR}$ in (U)LIRGs calculated using galaxy-wide SED fitting to obtain the dust mass are similarly low ($\delta_{\rm GDR}\sim 8-20$; \citealt{papa2010}). These values are significantly smaller than those observed in nearby normal, star-forming galaxies ($\delta_{\rm GDR}\sim 100-200$, \citealt{remy2014}). In addition, such low $\delta_{\rm GDR}$ values are hardly consistent with the strong far-IR molecular absorption observed in the far-IR \citep[e.g.][]{ga2015}, which favor MW-like values with molecular abundances consistent with chemical models \citep{ga2018}. Furthermore, spatially resolved observations of nuclear regions in (U)LIRGs also show MW-like high $\delta_{\rm GDR}$ \citep[e.g.][]{wilson2008}, instead of the extremely low values inferred in ULIRGs. 

This apparent contradiction is due to the degeneracy between the dust optical depth and the dust temperature. Emission from dust that is optically thick up to far-IR or sub-mm wavelengths \citep[e.g.][]{sakamoto2008, ga2015} naturally leads to excess FIR/sub-mm emission, and can dominate the emission from the cold dust component, leading to an overestimate of the dust mass. Distinguishing between these two cases is only possible with spatially resolved observations of the dust continuum in the long wavelength, optically thin regime.

\subsubsection{Mid-IR decomposition} \label{sssec:midIRdecomp}

Mid-IR spectroscopy offers a powerful way to determine the AGN fraction even in dust-obscured nuclei, as the blackbody emission from AGN-heated dust at $T \sim 1000 - 1500$ K at the inner boundary of the dust torus (the dust sublimation temperature) within $\sim 10$ pc of the AGN peaks in the mid-IR. For IRAS 20100-4516 and IRAS 03158+4227, we do not detect broad-line signatures in optical/near-IR/mid-IR spectra, suggesting that we are looking at the torus edge-on. We calculate the AGN fraction using the method developed in \citet{nardini2008}, based on the observation that the relative difference between mid-IR spectra of AGN and starburst dominated galaxies is the highest in the 
$5-8$ \mum regime. The $5-8$ \mum mid-IR spectra of starburst galaxies are fairly consistent \citep{brandl2006}, and are dominated by emission from Polycyclic Aromatic Hydrocarbon (PAH) features at 6.2 \mum and 7.7 $\mu$m, while those of AGN dominated galaxies show a smooth power-law continuum emission. PAH features are associated with star formation activity, as they exist in photodissociation regions and H{\sc II} regions, and are destroyed in the intense radiation fields surrounding an AGN \citep[e.g.][]{genzel1998, laurent2000}. The mid-IR spectra of ULIRGs can therefore be decomposed into their AGN and starburst components in the following manner 

\begin{equation} \label{eqn:fagn}
f^{\rm obs}_{\nu} = f^{\rm int}_{\nu}[(1 - \alpha_{6}) u_{\nu}^{\rm SB} + \alpha_{6} u_{\nu}^{\rm AGN} e^{- \tau(\lambda)}]
\end{equation}

where $f^{\rm obs}_{\nu}$ is the observed flux, $f^{\rm int}_{\nu}$ is the intrinsic flux density, $u_{\nu}^{\rm SB}$ and $u_{\nu}^{\rm AGN}$ are the SB and AGN templates, normalized at $6\mu$m, $\alpha_{6}$ is the fraction of the flux contributed by the AGN at 6$\mu$m, and the PAH features are automatically included in the SB template. We include the dust extinction in the form of $\tau(\lambda) \propto \lambda^{-1.75}$ \citep{draine2007}. There are additional features detected in the mid-IR spectra of ULIRGs, such as the 6 \mum water-ice absorption, and additional aliphatic hydrocarbon (HAC) features at 6.85 and 7.25 $\mu$m. We include the 6 \mum water-ice feature using the 15-K pure water-ice spectrum from the Leiden ice database\footnote{\url{ http://icedb.strw.leidenuniv.nl/}}, and model the 6.85 \mum and 7.25 \mum aliphatic hydrocarbon features as Gaussians. We apply these to the fitting by adding additional multiplicative terms to \autoref{eqn:fagn}.We create the starburst template by averaging the mid-IR spectra from five starburst-dominated ULIRGs, IRAS $10190+1322$, IRAS $12112+0305$, IRAS 17208$-$0014, IRAS 20414$-$1651 and IRAS 22491$-$1808 (all with $f_{\rm AGN} \lesssim 0.05$, using multi-band observations in the literature), and use a power-law template for the AGN contribution ($f_{\nu} \propto \lambda^{\gamma}$, with $\gamma \sim 0.7 - 1.5$). We leave the parameters for the ice features, $\alpha_{6}$, the optical depth at 6 \mum and the overall normalization as free parameters in the fitting. We extrapolate from the $\alpha_{6}$ to $f_{\rm AGN}$, the AGN contribution to the bolometric luminosity i.e $L_{\rm AGN} \sim f_{\rm AGN} L_{\rm IR}$, and thus the AGN luminosity $L_{\rm AGN}$ \citep[see][for more details]{nardini2008, nardini2010}.

\begin{figure*}
\includegraphics[width=0.49\textwidth]{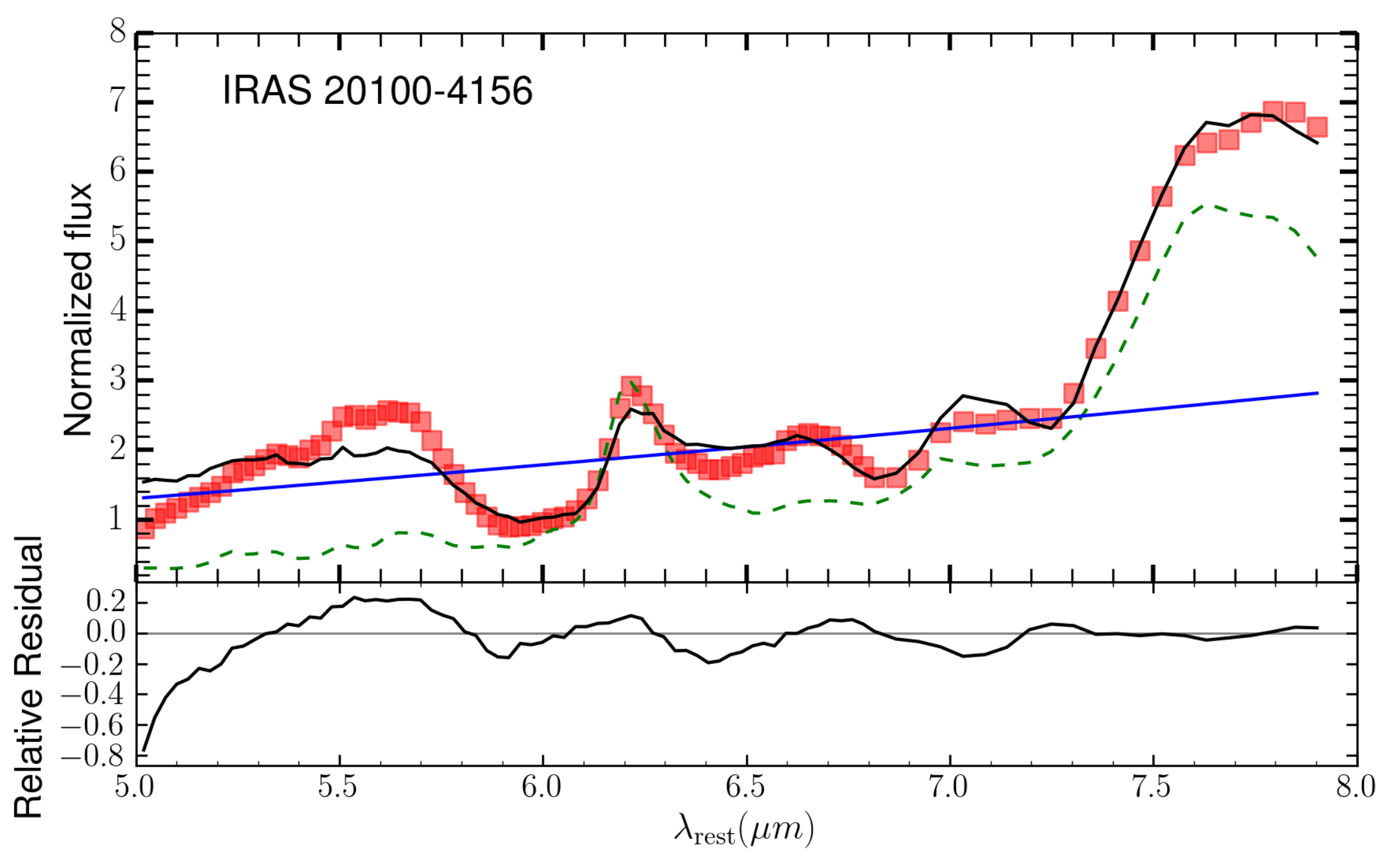}
\includegraphics[width=0.49\textwidth]{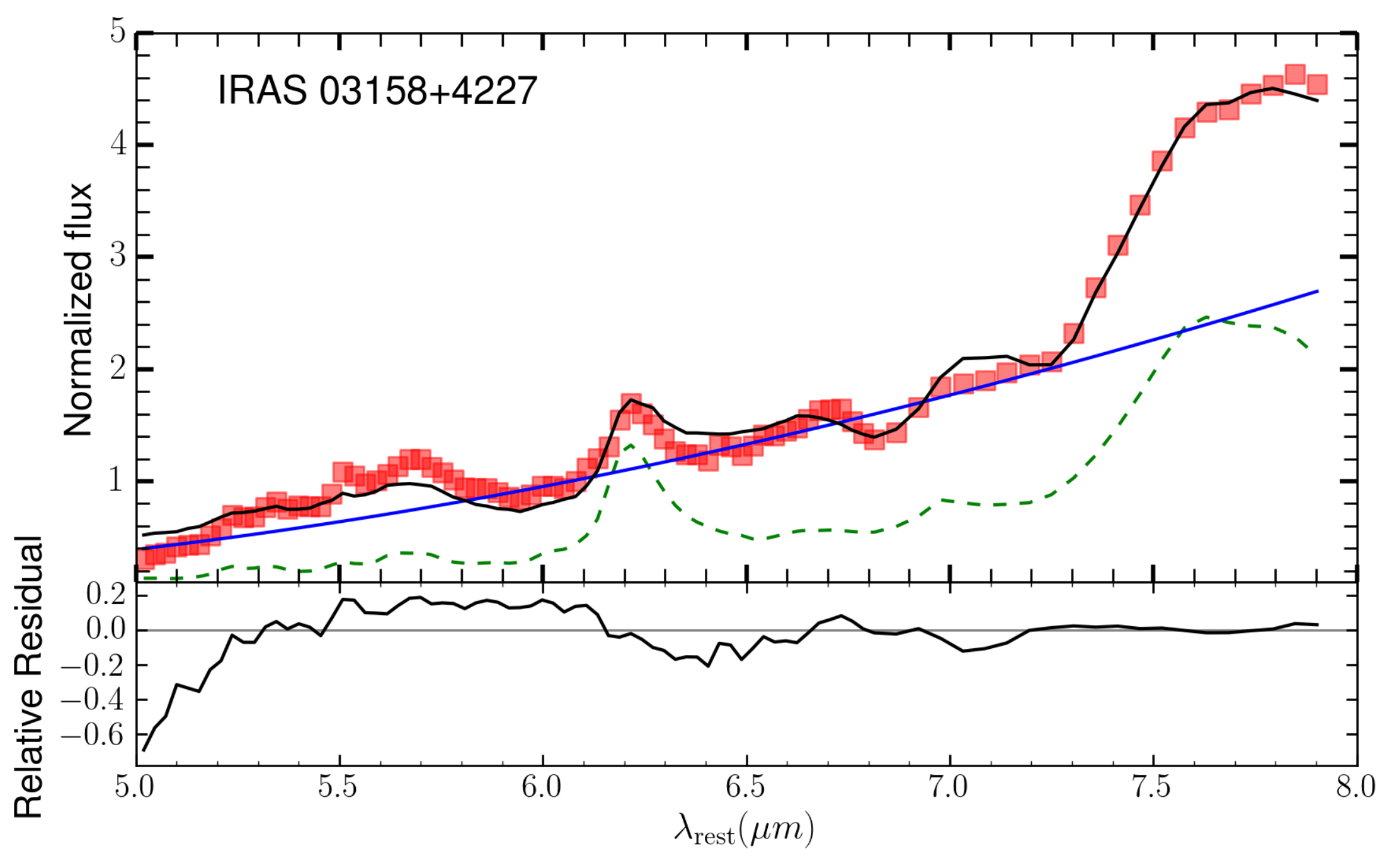}
\caption{Results of mid-IR spectral decomposition in IRAS 20100$-$4156 and IRAS 03158$+$4227. The dashed green line shows a starburst template, the red points show the observed IRS spectrum, and the blue line shows the AGN continuum emission, while the black line shows the weighted sum of the two components after incorporating water-ice features. The lower panels in each show the relative residuals at each point. }
\label{fig:midIRs_imp}
\end{figure*}

The best-fit mid-IR decompositions for IRAS 20100$-$4156 and IRAS 03158$+$4227 are shown in \autoref{fig:midIRs_imp}. We find that the best-fit model deviates from the observed mid-IR emission at 5-5.5 $\mu$m, and finds a shallower slope in that spectral range than is actually observed. The steeper slope observed is likely due to the extended $4-5$ \mum CO gas/ice absorption feature \citep[see][]{spoon2004}, which we have not included in our modeling. We find AGN fractions of $f_{\rm AGN} = 0.14 \pm 0.47 $ and $f_{\rm AGN} = 0.43 \pm 0.18$ for IRAS 20100$-$4156 and IRAS 03158+4227, respectively. These are consistent within the uncertainties with $f_{\rm AGN}$ found using SED fitting (\autoref{sssec:sedfitting}), and also consistent with the AGN fractions found previously for these objects; \citealt{nardini2009}, and those typically found in ULIRGs \citep{farrah2012}.

To enable comparison between IRAS 20100$-$4156, IRAS 03158+4227, and other molecular outflows detected using CO in ULIRGs, we have compiled all such sources from the literature \citep{cicone2014, garcia2015, veilleux2017}, and also performed a similar mid-IR decomposition for them. We find results generally consistent with previous AGN fractions for those sources included in the \citet{cicone2014} sample \footnote{Our values are within 0.1 dex for all such sources, except for IRAS 10565+2448, for which we find an $f_{\rm AGN}\sim 0.05$, while $f_{\rm AGN}\sim 0.17$ in the literature.}. The results of all these are shown in \autoref{tab:TableOutflow}.

\subsubsection{Comparison of AGN diagnostics against X-ray observations.}\label{sssec:xrays}

We compare $L_{\rm AGN}$ from mid-IR spectral decomposition to the previously obtained $L_{\rm AGN}$ from X-ray observations for our two sources. Primary hard X-ray spectra from AGN are characterized by an X-ray photon index $\Gamma$ (where the number of photons at energy E is given by $n(\rm E) \propto E^{-\Gamma}$), with the power-law cutoff energy determined by the column density $N_{\rm H}$, and increasing to higher photon energies at higher column densities. In addition to this, X-ray spectra from AGN-dominated ULIRGs also show a broad Fe K-$\alpha$ line at 6.4 keV, with equivalent width $\rm EW \sim 1$ keV. The soft X-ray emission ($\lesssim 1 $ keV) is often dominated by thermal emission from a hot plasma at $\sim 0.7$ keV from nuclear starburst activity. While primary hard-X ray emission from the AGN may be blocked at energies $<10$ keV in deeply obscured Compton thick sources ($N_{\rm H} \gtrsim 1.5 \times 10^{24}$ cm$^{-2}$), spectra up to 10 keV can be used to determine the power-law cutoff (which gives the column density), the Fe K-$\alpha$ line, and reflection components, depending on the geometry of the absorber. 

For IRAS 20100$-$4156, XMM-\emph{Newton} was used to observe the $0.5 - 10$ keV X-ray emission \citep{franceschini2003}. The emission was detected at modest significance, and the spectrum modeled using a combination of thermal plus a power-law emission (either absorbed or with a cut-off, both similarly good fits). The photon indices were fixed to $\Gamma \sim 1.1$ and $\Gamma \sim 1.7$, and both fits gave comparable values for $N_{\rm H} \sim 2 - 2.3 \times 10^{22}$ cm$^{-2}$, and intrinsic absorption-corrected $2-10$ keV luminosity of $L_{\rm 2 - 10\rm keV} = 1.6 \times 10^{42}$ ergs s$^{-1}$ i.e. IRAS 20100$-$4156 does not show AGN-dominated X-ray emission. AGN-host galaxies also show a relative FIR deficit (or X-ray excess) in the X-ray - $L_{\rm FIR}$ relation \citep{ranalli2003, franceschini2003}. IRAS 20100$-$4156 is consistent within the scatter in this relation, showing that the observed X-ray emission is predominantly from starburst activity.

However, we note that the obtained column density is lower than the column density obtained from modeling of far-IR OH lines \citep{ga2017}, which is $> 10^{24}$ cm$^{-2}$. It is then possible that IRAS 20100$-$4156 hosts an intrinsically X-ray weak and deeply obscured AGN, which is drowned out by the X-ray emission from less obscured circumnuclear starburst. Such X-ray weak AGN have been seen in local ULIRGs (including Mrk 231, \citealt{teng2014, teng2015}) and have been attributed to super-Eddington accretion onto the central SMBH \citep{vasudevan2009, lusso2010, lusso2012}. 

A similar lack of X-ray AGN signatures is found for IRAS 03158+4227. The Advanced Satellite for Cosmology and Astrophysics (ASCA) was used to observe the hard X-ray emission in IRAS 03158+4227 and detected a $2-10$ keV flux of $\sim (5.4 \pm 1) \times 10^{-13}$ ergs s$^{-1}$ cm$^{-2}$ \citep{risaliti2000}. This implies a $2-10$ keV luminosity of $L_{2-10 \rm keV} = (2.5 \pm 0.5) \times 10^{43}$ ergs s$^{-1}$, consistent with thermal emission, though \citet{risaliti2000} invoke high column densities of $N_{\rm H}> 10^{24}$ cm$^{-2}$ to explain the lack of X-ray AGN signatures. We also note that the ASCA field of view is $50'$, and the detected emission is likely contaminated by the interacting companion of IRAS 03158+4227, which is a Seyfert 1 galaxy \citep{meusinger2001}.

Overall we do not find X-ray signatures for AGN in either IRAS 20100$-$4156 and IRAS 03158+4227. We convert between $L_{2-10\rm keV}$ and the bolometric AGN luminosity using relations from \citet{marconi2004}, and find $L_{\rm AGN} \sim 2 \times 10^{10} L_{\odot}$ and $L_{\rm AGN} \sim 3 \times 10^{11} L_{\odot}$, which correspond to $f_{\rm AGN} \sim 0.005$ and $f_{\rm AGN} \sim 0.08$ for IRAS 20100$-$4156 and IRAS 03158$+$4227 respectively. The ratios of the $L_{2-10\rm keV}$ to the bolometric AGN luminosity are then 0.08\% and 0.4\% respectively. These are significantly lower than what is typically seen for Seyfert 1 galaxies (ratios between 2\% and 15\%, \citealt{elvis1994}) and are consistent with both sources being extremely under-luminous in X-ray emission, similar to that seen in other ULIRGs \citep{teng2015}, which has been attributed to higher Eddington ratios. Future hard X-ray observations of the two sources at energies $> 10$ keV are needed to confirm this hypothesis. Meanwhile, we treat the $L_{\rm AGN}$ from X-ray observations as lower limits on the AGN luminosity.

\subsubsection{Other AGN diagnostics} \label{sssec:othersmidIR}

We also mention other commonly used optical/near-IR/mid-IR AGN diagnostics for completeness. These include high-ionization mid-IR lines - e.g. [Ne {\sc V}] 14.3 \mum and [Ne {\sc VI}] 7.6 \mum - which are only detected in AGN-host sources, and the optical lines [O III]/H$\beta$, whose ratio is used to characterize line ionization and the hardness of the radiation field. 

Neither IRAS 20100$-$4156 or IRAS 03158$+$4227 display the high-ionization lines which are most commonly seen in AGN host galaxies, though they display the lower-energy [Ne II] and [Ne III] lines \citep{farrah2007}. The ratio of the two probes the hardness of the radiation field, with starburst dominated galaxies showing a line ratio of $-1.3 \lesssim \rm log([Ne III]/[Ne II]) \lesssim 0.1$ in a sample of GOALS galaxies, though highly starbursting galaxies can show high ratios up to $\rm log ([NeIII]/[NeII]) > -0.2$; \citealt{inami2013}). We find log([Ne III]/[Ne II]) $\sim -0.4$ and $\sim -0.8$ for IRAS 20100$-$4156 and IRAS 03158+4227, respectively, which are consistent with the starburst range. Similarly, IRAS 20100$-$4156 and IRAS 03158+4227 show [O III]/H$\beta \sim 1.2$ and [O III]/H$\beta \sim 2.3$, respectively \citep{staveley1992, meusinger2001}, which are too low for AGN, but are consistent with LINER ratios based on the BPT diagnostic diagram \citep{kewley2006}. Near-IR spectroscopy of the Pa-$\alpha$ and Br$\gamma$ lines for IRAS 20100$-$4156 also does not show broad pedestals in the spectral lines which would be evidence for emission from a broad line region (BLR). 

Overall, these diagnostics are consistent with the significant dust-obscuration observed in the two sources, and we therefore cannot comment on the AGN luminosity based on any of these. 

\subsubsection{Uncertainties in AGN fraction}\label{ssec:caveats}

IRAS 20100$-$4156 and IRAS 03158+4227 are extremely dust-obscured sources, with high silicate optical depths \citep{spoon2013}, and high nuclear column densities, as evidenced by the deep absorption in far-IR OH lines \citep{ga2017}. This extreme obscuration renders most line diagnostics ineffective. This high dust-obscuration also impacts the mid-IR decomposition, which is no longer applicable if the hot dust emission is completely obscured by cooler surrounding dust, as the warmer photons will be re-emitted at longer wavelengths. Another caveat for mid-IR decomposition is that the starburst template used for decomposition is derived using a combination of mid-IR spectra from starburst-dominated ULIRGs, which could always include some small AGN component in principle, possibly biasing the method towards finding higher starburst fractions. We note, however, that IRAS 08572+3915, an AGN-dominated highly obscured ULIRG \citep{efstathiou2014}, is correctly identified as a highly AGN-dominated galaxy based on mid-IR decomposition, despite showing a deeper silicate optical depth (also no [Ne {\sc V}] emission), and a geometry consistent with an edge on torus, similar to the geometry for IRAS 20100$-$4156 and IRAS 03158+4227. This strongly suggests that a significant fraction of mid-IR photons can escape through optically thin lines of sight, despite high dust opacities. This scenario has been supported by simulations \citep[e.g.][]{roth2012, novak2012, krumholz2013,skinner2015}, although the results are sensitive to the numerical methods in question \citep{davis2014, tsang2015}. Overall, we do not find any evidence that the mid-IR decomposition is finding artificially low AGN fractions for our targets, though we cannot rule out more luminous and deeply buried AGN with their dust torus completely obscured by cooler molecular gas. 

\begin{table*}
\setlength{\tabcolsep}{1pt}
\renewcommand{\arraystretch}{1.1}
\begin{center}
\caption{Properties of outflow wings (see \autoref{sssec:iras20100_wings}, \autoref{sssec:iras03158_wings} for details)}
\begin{tabular}{lccccccc}
\tableline
\tableline
Source 		&
Core	&
\multicolumn{2}{c}{red wing} 	&
\multicolumn{2}{c}{blue wing} 	& 
size \\ \toprule
					& {$S_{\rm CO}$} 	& {Velocity range} 	& $S_{\rm CO}$  	& {Velocity range} 	& $S_{\rm CO}$  & {R} 	& \\
					& ({Jy km s$^{-1}$}) 	& ({km s$^{-1}$})	 	& ({Jy km s$^{-1}$}) 	& ({km s$^{-1}$}) & ({Jy km s$^{-1}$})  	& ({kpc}) & \\ \midrule
IRAS 20100$-$4156 		& ($11.2 \pm 0.3$) 	& (440,1650) 		& ($0.54 \pm 0.03$) & (-1169, -466) 	& ($0.58 \pm 0.03$) 	& $1.0 \pm 0.2$ 	& \\
IRAS 03158$+$4227 	& ($8.4 \pm 0.7$) 	& (400,1750) 		& ($0.80 \pm 0.22$) & (-944, -413) 	& ($0.31 \pm 0.15$) 	& $1.9 \pm 1.3 $ 	& \\ 
\tableline
\end{tabular}
\label{tab:wp}
\end{center}
\end{table*}

\setlength{\tabcolsep}{1pt}
\renewcommand{\arraystretch}{1.1}
\begin{table*}
\begin{center}
\caption{All properties of outflow sources.}
\begin{tabular}{lccccchcccccc}
\tableline
Name	&  $z$	& $M_{\rm gas}$\tablenotemark{a} 	& log$_{10}(L_{\rm IR})$\tablenotemark{b}	& $f_{\rm AGN}$\tablenotemark{c} 	& $L_{\rm AGN}$ \tablenotemark{d}	& SFR 					& $\dot M_{\rm OF}$ 	& $v_{\rm OF}^{\rm AVG}$ 		& log$_{10}(\tau_{\rm dep})$ 	& $\dot P_{\rm OF}$/($L_{\rm AGN}/c$) 	& log$_{10}(\dot E_{\rm OF}$)  & Ref \\
		& 	 	& ($M_{\odot}$) 	& ($L_{\odot}$) 			& 				& (ergs s$^{-1}$)		&($M_{\odot}$ yr$^{-1}$) 		& ($M_{\odot}$ yr$^{-1}$) 	& (\kms) 						& (yrs)						 &								& (ergs s$^{-1}$) &   \\ 
\tableline
IRAS 20100$-$4156		& $ 0.129 $ & $ 9.85 $ 	& $ 12.63 $	& $ 0.14 $& $ 45.35 $ & $ 347 $ & $ 672 $ & $ 929 $ & $ 7.04 $ & $ 53 $ & $ 44.26 $ & (1)			\\
IRAS 03158+4227		& $ 0.134 $ & $ 9.75 $ 	& $ 12.6 $ 	& $ 0.43$ & $ 45.82 $ & $ 306 $ & $ 350 $ & $ 876 $ & $ 7.11 $ & $ 9 $ & $ 43.93 $ & (1)\\
\tableline
Mrk 231				& $ 0.042 $ & $ 9.73 $ 	& $ 12.49 $ 	& $ 0.42 $& $ 45.70 $ 	& $ 196 $ 	& $ 352 $ & $ 700 $ & $ 7.19 $ & $ 9 $ & $ 43.73 $ & (2)\\
IRAS 08572+3915		& $ 0.058 $ & $ 9.18 $ 	& $ 12.10 $ 	& $ 0.91 $& $ 45.64 $ 	& $ 80 $ 	& $ 406 $ & $ 800 $ & $ 6.57 $ & $ 14 $ & $ 43.91 $ & (3)\\
IRAS F10565+24489 	& $ 0.043 $ & $ 9.90 $ 	& $ 12.00 $ 	& $ 0.05 $	& $ 44.28 $ 	& $ 84 $ 	& $ 98 $ & $ 450 $ & $ 7.91 $ & $ 43 $ & $ 42.79 $ & (3)\\
IRAS 23365+3604		& $ 0.064 $ & $ 9.93 $ 	& $ 12.17 $ 	& $ 0.05 $	& $ 44.46 $ 	& $ 111 $ 	& $ 55 $ & $ 450 $ & $ 8.19 $ & $ 16 $ & $ 42.54 $ & (3)\\
Mrk 273				& $ 0.038 $ & $ 9.70 $ 	& $ 12.13 $ 	& $ 0.07 $	& $ 44.56 $ 	& $ 107 $ 	& $ 200 $ & $ 620 $ & $ 7.40 $ & $ 65 $ & $ 43.38 $ & (3)\\
IRAS 17208$-$0014\tablenotemark{e}	& $ 0.042 $ 	& $ 9.50 $ 	& $ 12.38 $ & $ 0.05 $& $ 43.97 $ 	& $ 206 $ 	& $ 163 $ & $ 400 $ & $ 7.29 $ & $ 131 $ & $ 42.91 $ & (4)\\
IRAS 11119+3257\tablenotemark{e}  	& $ 0.190 $ 	& $ 9.95 $ 	& $ 12.65 $ & $ 0.48 $& $ 45.91 $ 	& $ 242 $ 	& $ 153 $ & $ 600 $ & $ 7.67 $ & $ 2 $ & $ 43.18 $ & (5) \\ 
\tableline

\label{tab:TableOutflow}
\end{tabular}
\end{center}
\tablenotetext{a}{Calculated assuming an \aco $= 0.8$ \acou.}
\tablenotetext{b}{Calculated using the prescription from \citet{sanders1996}}
\tablenotetext{c}{Calculated using mid-IR decomposition}
\tablenotetext{d}{$L_{\rm AGN} = f_{\rm AGN} L_{\rm IR}$. }
\tablenotetext{e}{For these two ULIRGs we have estimated the average outflow velocity using visual inspection of the published spectra.}
\tablerefs{(1) This work (2) \citet{cicone2012} (3) \citealt{cicone2014} (4) \citet{garcia2015} (5) \citet{veilleux2017}}

\end{table*}

\setlength{\tabcolsep}{3pt}
\begin{table*}[]
\begin{center}

	\caption{\textsc{SED-fitting results and other physical parameters (\autoref{sssec:sedfitting})}}
	\begin{tabular}{lcccccccccc}
  \hline 
	Source	& $z$ 	& 	$f_{\rm AGN}$	&	$L_{\rm dust}$	&   $U_{\rm min}$ 	& $M_{\rm dust}$	& 	$M_{\rm gas}$	& $M_{*}$ 	& SFR	& sSFR	&  $\delta_{\rm GDR}$\\
			& 		& 				&($10^{11} L_{\odot}$)	&	& (10$^{8}$ $M_{\odot}$) 	& (10$^{9}$ $M_{\odot}$)	& (10$^{11}$ $M_{\odot}$)	&( $M_{\odot}$ yr$^{-1}$)	& (Gyr$^{-1}$)&\\
	\hline
    IRAS 20100$-$4156 	&$0.129$ 	&$0.2\pm 0.1$	& 	$2.9 \pm 0.2$ 	&	35	& 	$4.2 \pm 0.2 $   &  	7.8	& $ 5.5 \pm 0.3 $  	& $198\pm 10$ 	& 3.2 	& 16	\\
	IRAS 03158$+$4227  	&$0.134$ 	&$0.3\pm 0.1$	& 	$2.9 \pm 0.2$ 	&	29	& 	$5.0 \pm 0.5$  	&  	6.4	& $ 18.6 \pm 0.9$   	& $244 \pm 12$ 	& 1.4		& 13	 	\\ 
	\hline \noalign {\smallskip}
	\end{tabular}
    \end{center}

	\tablecomments{The gas masses were calculated using $\alpha_{\rm CO} =  0.8M_{\odot} $(K km s$^{-1}$ pc$^{2})^{-1}$, based on CO($J = 1 \to 0$) observations.} 
	\label{tab:sedfits} 
\end{table*}

\section{Discussion}\label{sec:discussion}

Here we discuss the outflows in IRAS 20100$-$4156 and IRAS 03158$+$4227 in the context of other known molecular outflows. We compile all published observations of molecular gas outflows using CO($1-0$) in the local universe (\autoref{tab:TableOutflow}). We restrict this sample to ULIRGs, as feedback in lower-mass galaxies ($M_{*} \sim 10^{10} - 10^{11} M_{\odot}$) often occurs in `radio' mode via relativistic jets. Such outflows have much smaller mass outflow rates ($\dot M \lesssim 50 M_{\odot}$ yr$^{-1}$) and may quench star formation by injecting turbulence and decreasing the star formation efficiency \citep[e.g.][]{alatalo2015a} rather than sweeping out the molecular gas and dust at high rates. Given the physically distinct feedback mechanism in these systems, we do not include them in our analysis. Furthermore, the method of mid-IR decomposition has been calibrated for use in ULIRGs \citep{nardini2010,nardini2011}. Most of the ULIRGs in our sample also show signatures of outflowing molecular gas in \emph{Herschel} far-IR OH spectra (see \autoref{tab:TableOutflow} and references within), and are thus drawn from a similar parent sample as IRAS 20100$-$4156 and IRAS 03158+4227. Since more IR-luminous galaxies typically become more AGN-dominated, including sources with IR luminosities varying over 2 orders of magnitudes can potentially introduce a bias towards finding a correlation between outflow properties and AGN luminosities. This is avoided by selecting galaxies at comparable IR luminosities. 

We determine the outflow energy flux and momentum for this sample of ULIRGs as described in \autoref{ssec:outflows}. We determine the AGN fraction $f_{\rm AGN}$ for these sources using mid-IR decomposition, as described in \autoref{ssec:agnfrac}, and calculate the IR luminosity as described in \citet{sanders1996}. This ensures that IR luminosities are compared consistently between our targets and the rest of the ULIRG sample. The AGN and nuclear starburst luminosities are defined as $L_{\rm AGN} \sim f_{\rm AGN} L_{\rm IR}$ and $L_{*} \sim (1-f_{\rm AGN}) L_{\rm IR}$, respectively. Together with IRAS 20100$-$4156 and IRAS 03158+4227, we also highlight outflow properties for IRAS 08572+3915 and Mrk 231, which host comparably powerful outflows, while featuring dust obscuration properties (based on the silicate optical depth $\tau_{\rm 9.7 \mu m}$ as defined by \citealt{spoon2007}) which bracket the two sources of interest, IRAS 03158+4227 and IRAS 20100$-$4156.

\subsection{Outflow energetics} 

Galaxy-wide molecular outflows can be driven by AGN, nuclear starbursts, or a combination thereof, all of which will result in different outflow energy and momentum fluxes. AGN-driven outflows can have two phases: an energy-conserving phase, where the hot shocked bubble expands on timescales significantly shorter than its radiative cooling timescale and outflow energy is conserved, or a momentum-conserving phase, wherein the expansion and cooling timescales are comparable and the momentum is conserved \citep[e.g.][]{zubovas2012}. Energy conserving outflows are created when AGN-driven energetic nuclear winds ($v_{\rm in} \sim 0.1-0.3c$) collide with the ISM, resulting in shocked hot gas which cools inefficiently. The resulting outflow is nearly adiabatic and can produce significant momentum boosts\footnote{see \citealt{ga2017}, section 7.2 for a derivation} 
$\dot P \sim \beta (L_{\rm AGN}/c) v_{\rm in}/v_{\rm OF} \sim 10 L_{\rm AGN}/c $ for $\beta \sim 0.5$, $v_{\rm in } \sim 0.1c$ and $v_{\rm OF} \sim 1500$ \kms, where $\beta$ is the fraction of the energy converted to bulk motion of the ISM (derived to be $\beta \sim 0.5$ for an expanding adiabatic bubble, while the rest of the energy goes into the thermal pressure of the shocked bubble; \citealt{weaver1977, fg2012}). This model is observationally supported by detections of high momentum boosts in molecular gas outflows, which imply that nearly all outflows in ULIRGs pass through at least a partially energy-conserving phase, during which the momentum boost is imprinted \citep{ga2017}. Furthermore, relativistic ionized nuclear winds which drive the outflows have been detected using X-ray spectroscopy \citep{tombesi2015, feruglio2015, nardini2015}, and energy has been shown to be conserved between the relativistic and molecular phases of the outflow \citep{feruglio2015,veilleux2017}. Theoretically, the mechanical luminosity imparted to the nuclear wind is $\dot E \sim \dot M_{\rm in} v_{\rm in}^{2}/2 \sim (L_{\rm AGN}/c) v_{\rm in}/2 \sim 5\% L_{\rm AGN}$ for $v_{\rm in} \sim 0.1 c$ \citep[see][for a derivation]{king2015}. Of this, $\sim 20\%-50\%$ can be translated into bulk motion of the shocked ISM layer \citep{fg2012,richings2017b}. 

On the other hand, feedback from nuclear starburst activity takes place via supernovae, stellar winds, and H{\sc II} regions, and is expected to produce more modest momentum boosts; for example, a 40 Myr old starburst produces a momentum boost of $\dot P \sim 3.5 L_{*}/c$ \citep{leitherer1999, veilleux2005}. This expression includes both the momentum deposition by SNe and stellar winds at a level $\sim 2.5 L_{*}/c$, and a radiation pressure term $L_{*}/c$ under the assumption of a single photon scattering event. In the case of radiation pressure on optically thick dust (column densities $\geq 10^{23}$ cm$^{-2}$), multiple photon scatterings can lead to a higher momentum boost, $\dot P \sim \tau_{\rm IR} (L/c)$, where $L$ is either the AGN or the starburst luminosity \citep{thompson2015}. However, this cannot lead to arbitrarily high momentum boosts even in nuclei with large IR optical depths, and the momentum boost is limited to $\dot P\lesssim 5L_{\rm IR}/c $, as radiative transfer models for dusty nuclei find that photons can escape through optically thin lines of sight despite high dust opacities \citep[e.g.][]{roth2012, novak2012, krumholz2013,skinner2015}, although the results are sensitive to the numerical methods in question \citep{davis2014, tsang2015}. The maximum mechanical energy that can be transferred to the outflow for a 40 Myr old starburst is $\dot E = 7 \times 10^{41}$ SFR/($M_{\odot}$ yr$^{-1})$ ergs s$^{-1}$, which is equivalent to $\dot E \sim 1.8\% L_{*}$, of which $\sim 25\%$ can be efficiently transferred to bulk motion of the gas. Comparing observed outflow energy and momenta (\autoref{tab:TableOutflow}) to different theoretical predictions therefore allows us to constrain the outflow driving mechanism. 

\subsubsection{Momentum}

\begin{figure*}
\includegraphics[width=\textwidth]{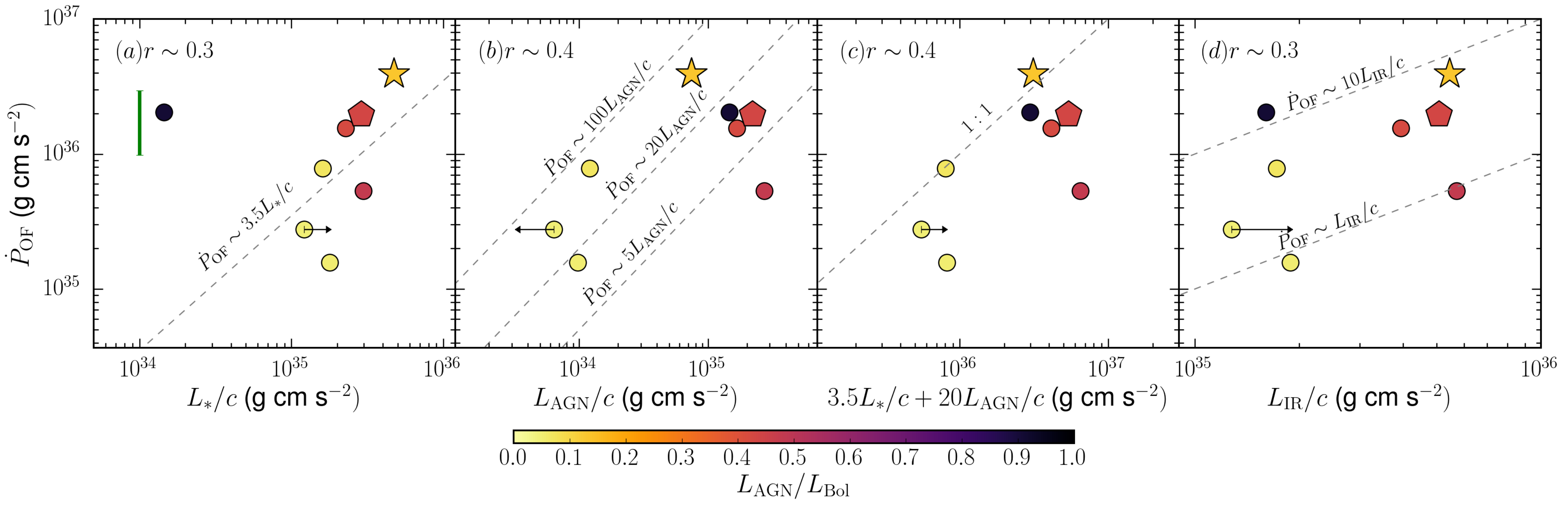}
\caption{Inferred outflow momentum flux ($\dot P_{\rm OF}$) as compared to (a) $L_{\rm *}/c$ (b) $L_{\rm AGN}/c$ (c) the predicted sum of the AGN and starburst momentum fluxes, $\dot P_{\rm sum}$, and (d) $L_{\rm IR}/c$. The star and pentagon show IRAS 20100$-$4156 and IRAS 03158$+$4227, respectively, while filled circles represent other ULIRGs for which molecular outflows have been observed in CO. Dashed lines show the expected coupling fraction between the momentum fluxes from AGN/starburst and the outflow momentum flux. The color of the points represents the $f_{\rm AGN}$, with darker points being more AGN-dominated galaxies. The green bar shows the representative uncertainty in the outflow momentum flux.}
\label{fig:momout} 
\end{figure*}

Here we compare the inferred outflow momentum fluxes ($\dot P_{\rm OF}$) to the expected momentum injection due to starburst and/or AGN activity i.e. $L_{\rm AGN}/c$, $L_{*}/c$, $L_{\rm IR}/c$, and the predicted momentum injected from the sum of both (to first order, $\dot P_{\rm sum} \sim 20 L_{\rm AGN}/c + 3.5 L_{*}/c$) for our sample of ULIRGs (\autoref{fig:momout}). 

We find that starburst activity alone is sufficient to provide the observed momentum boosts only in the weakest three outflows. We also find that nearly all the outflows show significant momentum boosts relative to the $L_{\rm AGN}$, $\dot P_{\rm OF} \sim (5-50) L_{\rm AGN}/c$. While these results are consistent with previous molecular gas outflow observations \citep[e.g.][]{cicone2014, ga2017}, we also find significant differences. We find that high momentum boosts up to $\dot P_{\rm OF} \sim 10 L_{\rm IR}/c$ are not restricted to ULIRGs with $f_{\rm AGN} \gtrsim 0.5$, but are seen in ULIRGs with a range of $f_{\rm AGN} \sim 0.2 - 0.9 $. We detect no significant correlation between the outflow momentum flux and $L_{*}/c$ and $L_{\rm IR}/c$ (Spearman correlation coefficient $r(\dot P_{\rm OF}, L_{\rm IR}/c) \sim r(\dot P_{\rm OF}, L_{\rm *}/c) \sim 0.3$ ; p-value $\sim 0.5$). We find a similar lack of correlation between the inferred outflow momentum flux and $L_{\rm AGN}/c$ (Spearman correlation coefficient $r(\dot P_{\rm OF}, L_{\rm IR}/c) \sim 0.4$ ; p-value $\sim 0.3$) and $\dot P_{\rm sum}$ (Spearman correlation coefficient $r(\dot P_{\rm OF}, \dot P_{\rm sum}) \sim 0.4$; p-value $\sim 0.4$)\footnote{The Spearman coefficient measures the monotonicity of the correlation between two quantities, while a Pearson correlation measures the linearity of the correlation between two. We did not recover significant correlations using either diagnostic.}). \footnote{IRAS 08572+3915 is the only AGN-dominated ULIRG with a relatively low total $L_{\rm IR}$. However, it has been suggested that due to the extreme dust obscuration, and the torus axis being perpendicular to the line of sight, its IR luminosity could be underestimated by as much as a factor of $5$ \citep{efstathiou2014}. Correcting for this factor would marginally improve the correlations described above. }

The observed momentum boost for IRAS 03158$+$4227 ($\dot P \sim 9 L_{\rm AGN}/c$) is consistent with the boosts predicted for AGN-driven outflows. For IRAS 20100$-$4156, on the other hand, we observe a high momentum boost ($\dot P \sim 53 L_{\rm AGN}/c$). The momentum boost is higher than expected in starburst-dominated outflow sources ($\dot P \sim 10 L_{*}/c$ for IRAS 20100$-$4156, while $\dot P \sim 2-4 \times L_{\rm *}/c$ for starburst galaxies, \citealt{leitherer1999,veilleux2005}). Such high momentum boosts can in principle be generated in deeply dust-obscured starburst nuclei ($\tau_{\rm IR} \gg 1$) with high surface densities of star formation ($\Sigma_{\rm SFR}$), such that the star formation is Eddington-limited by radiation pressure on dust ($\Sigma_{\rm SFR} \sim 3000 M_{\odot}$ yr$^{-1}$ kpc$^{-2}$, \citealt{murray2005, thompson2005}). However, \citet{roth2012} find that in such optically thick nuclei photons preferentially escape along optically thin lines-of-sight and limit the momentum boost to $\sim 5 L_{\rm IR}/c$. Furthermore, the high maximum outflow velocity in IRAS 20100$-$4156 is also unusual for starburst driven outflows, although not impossible. Previous studies have consistently found significant correlations between the maximum outflow velocities in molecular gas outflows and $L_{\rm AGN}$ \citep{sturm2011,spoon2013,veilleux2013,stone2016}, suggesting that AGN are responsible for the fastest-moving components of the outflow. Compact, Eddington-limited starburst activity has also been suggested to drive the high-velocity molecular outflow ($\sim 1000$ \kms) in a $z\sim 0.7$ galaxy \citep{geach2014}, although the host galaxy is much less massive. While observations of outflows in starburst galaxies show that their maximum outflow velocities are strongly correlated with $\Sigma_{\rm SFR}$, these probe ionized gas outflows, and the star formation surface densities are far below the Eddington limit \citep{heckman2015,heckman2016}. For IRAS 20100$-$4156, assuming a radius of $\lesssim 0.2$ kpc for the nuclear starburst and calculating the nuclear SFR $\sim 10^{-10} (L_{*}/L_{\odot}) (M_{\odot} \rm yr^{-1})$, we find a surface density of star formation $\Sigma_{\rm SFR} \gtrsim 3000 M_{\odot}$ yr$^{-1}$ kpc$^{-2}$, which is above the Eddington limit. Overall, the outflow in IRAS 20100$-$4156 could in principle be driven purely by the starburst. 

However, we consider this scenario unlikely based on the high outflow velocity observed. We instead favor the scenario that the outflow in IRAS 20100$-$4156 is driven by a combination of AGN and starburst activity, as neither AGN or the starburst feedback alone can provide the observed momentum boost, which we reiterate is a conservative estimate. 

\subsubsection{Energy}

\begin{figure*}
\includegraphics[width=\textwidth]{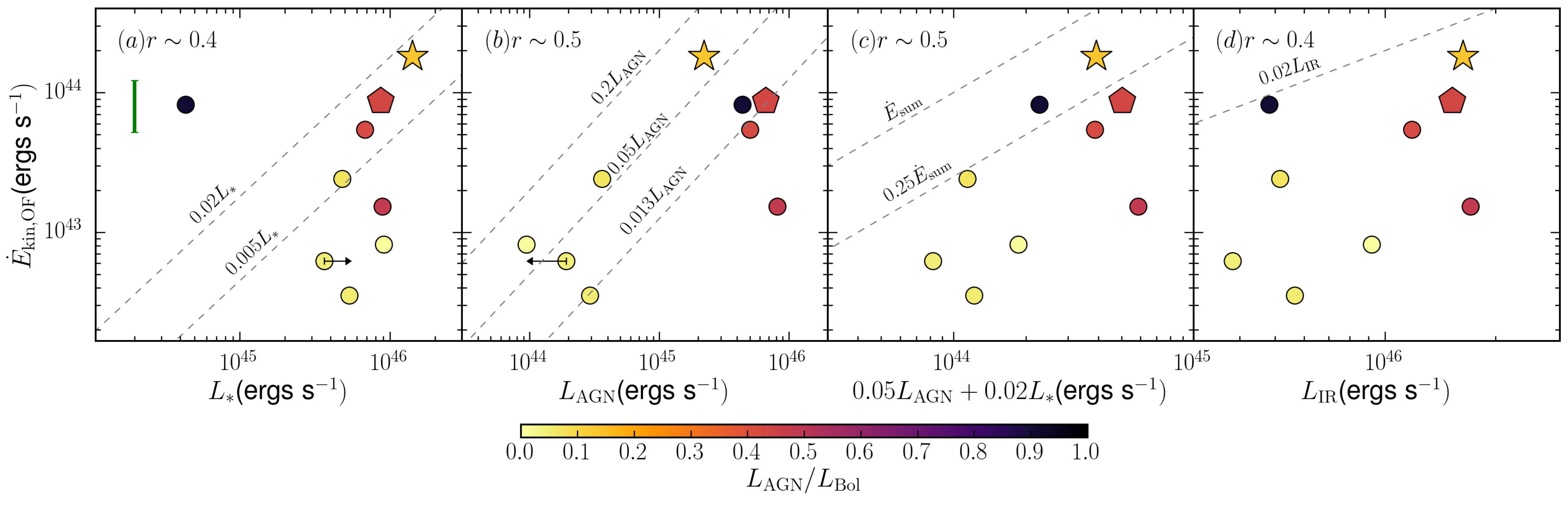}
\caption{Inferred outflow kinetic energy flux ($\dot E_{\rm kin, OF}$) compared to (a) $L_{*}$ (b) $L_{\rm AGN}$, (c) the predicted sum of AGN and starburst energies $\dot E_{\rm sum}$, and (d) $L_{\rm IR}$ respectively for the ULIRG sample. The conventions are the same as in \autoref{fig:momout}, and the dashed lines show the expected coupling fraction between AGN/starburst luminosities and the outflow kinetic energy flux. IRAS 20100$-$4156 would need to have an unrealistic coupling efficiency of $\gtrsim$ 8\% with the AGN luminosity, suggesting that both AGN and starburst activity need to play a role in driving the outflow in IRAS 20100$-$4156. The green errorbar shows the representative uncertainty in the outflow energy fluxes for the sample. }
\label{fig:eout}
\end{figure*}

We compare the outflow energy flux ($\dot E_{\rm OF}$) for all ULIRGs in our sample to $L_{\rm AGN}$, $L_{\rm *}$, $L_{\rm IR}$, and the predicted energy flux from a combination of AGN and starburst activity ($\dot E_{\rm sum} = 0.05 L_{\rm AGN} + 0.02 L_{*}$, \autoref{fig:eout}). We find that starburst activity can power the most massive outflows only if all the mechanical luminosity from supernovae and stellar winds is converted into bulk motion of the gas with 100\% efficiency, which is unrealistic. However, weaker outflows can be powered by stellar feedback even when realistic efficiencies are taken into account. Similarly, the outflows can be powered purely by AGN (i.e show energy fluxes $\sim 1-3\% L_{\rm AGN}$) only in ULIRGs with $f_{\rm AGN} \gtrsim 0.4$, but not in starburst dominated galaxies. We do not find that the outflow energy fluxes increase significantly with $L_{\rm AGN}$ and with $\dot E_{\rm sum}$ (Spearman correlation coefficients $r(\dot E_{\rm OF}, L_{\rm AGN}) \sim 0.5$, $r(\dot E_{\rm OF}, \dot E_{\rm sum}) \sim 0.5$ ; p-values $\sim 0.3$ and $\sim 0.2$, respectively). We find a similar lack of correlation with $L_{*}$ and $L_{\rm IR}$ (Spearman correlation coefficients $r(\dot E_{\rm OF}, L_{\rm *}) \sim 0.4$, $r(\dot E_{\rm OF}, L_{\rm IR}) \sim 0.4$ ; p-value $\sim 0.4 $ for both). We find that powerful outflows are detected in ULIRGs with $f_{\rm AGN}$ with a wide range of $\sim 0.2 -0.9$. 

We find that IRAS 03158$+$4227 shows an outflow energy flux of $\sim 1\% L_{\rm AGN}$, which is close to the expected scaling. The outflow in IRAS 20100$-$4156, however, shows an outflow energy flux of $\sim 8\% L_{\rm AGN}$, which is significantly higher than predicted from AGN activity alone. We therefore conclude that it is driven by a combination of the two, which is consistent with findings in the previous section.

\subsection{What drives the most powerful outflows?}\label{ssec:nowrongchoices}

We have compared the outflow energy and momentum fluxes in IRAS 20100$-$4156 and IRAS 03158+4227 against theoretical predictions and outflows in other ULIRGs, drawn from the same parent ULIRG sample and with similar observations of CO($1-0$). Despite the relatively small range in $L_{\rm IR}$, our sample demonstrates a range in $\dot P_{\rm OF}$ and $\dot E_{\rm OF, kin}$ varying over two orders of magnitude, and can therefore be used to test for correlations with the AGN and starburst luminosities. For the complete sample, we find no significant correlation between the molecular outflow properties and $L_{\rm AGN}$, $L_{\rm *}$, or a combination of the two. We also find that the strongest outflows span a large range in $f_{\rm AGN} = 0.2 - 0.9$, despite having very similar IR luminosities i.e. $f_{\rm AGN}$ is a poor proxy for the strength of feedback \citep[also see][]{farrah2012}. We also find that starburst activity can play an important role even in the most massive outflows, as demonstrated by IRAS 20100$-$4156. The lack of dependence of outflow properties on $f_{\rm AGN}$ suggests that previously seen correlations between outflow properties and AGN luminosities may be partly driven by the bias that ULIRGs statistically show higher $f_{\rm AGN}$ at higher IR luminosities \citep{nardini2010,alonso2012, alonso2013}, as compared to smaller galaxies.

\subsection{OHM as a possible new tracer for outflows}\label{ssec:ohm}

\begin{figure*}
	\centering
	\includegraphics[width=\textwidth]{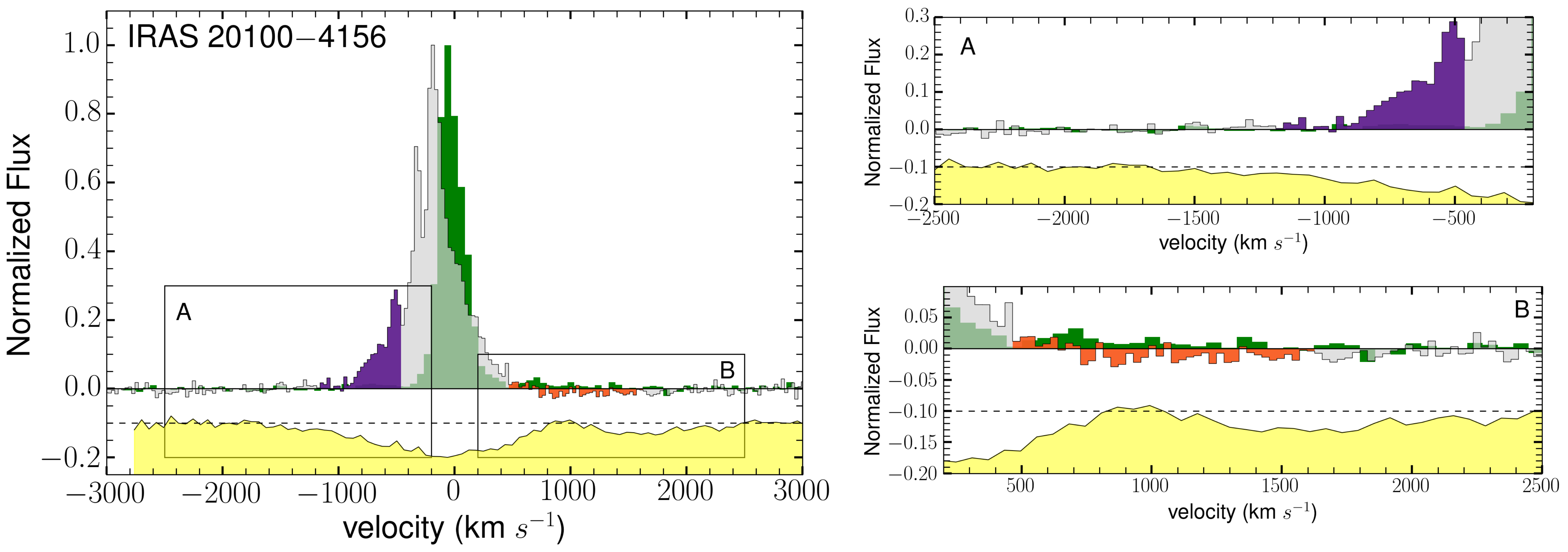}
	\caption{Overlaid spectra for CO($1-0$) emission (green), OH 119 \mum P-Cygni line profile (yellow, normalized to a zero of $-0.1$), and OH 1.667 GHz emission (grey) from IRAS 20100$-$4156. The purple and orange regions show the red and blue outflow wings respectively. The A and B panels show the zoomed in regions around the red and blue outflow wings. We detect only the blue outflow wing in OHM emission, while the red wing shows tentative evidence for absorption.}
	\label{fig:iras20100infall}
\end{figure*}

\begin{figure}
	\centering
	\includegraphics[width=0.4\textwidth]{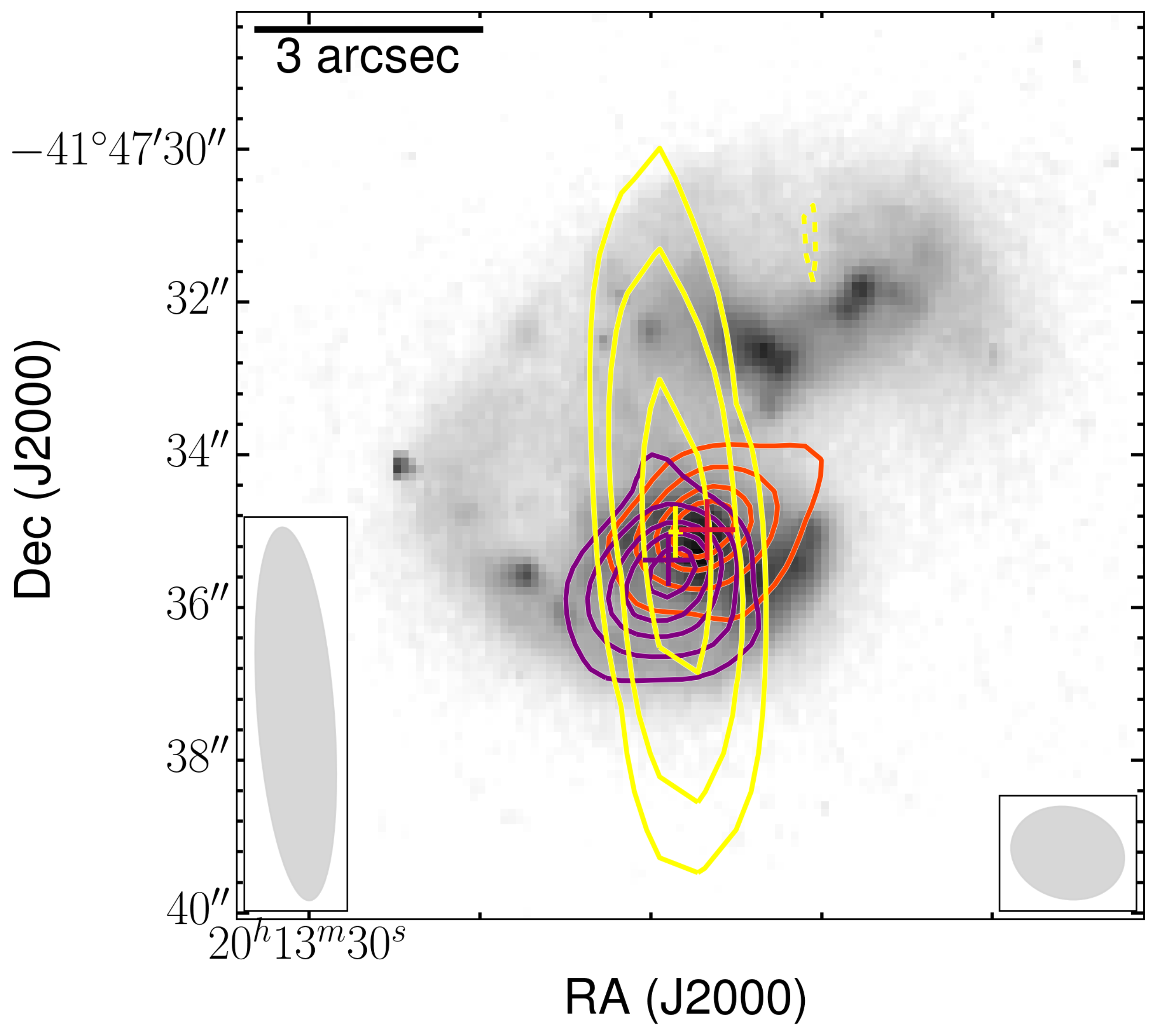}
	\caption{Spatial extent of the OHM emission as compared to the red and blue outflow wings seen in CO($1-0$) in IRAS 20100$-$4156. Emission is shown as $\pm 4,5,6\sigma$ significance contours for the outflow wings, and $\pm 10,20,30,40\sigma$ contours for OHM emission. Colored crosses show the peak pixel and their lengths represent the $3\sigma$ positional uncertainty for each of OHM (yellow), red, and blue outflow wings. The OHM emission is nearly co-spatial with the blueshifted outflow wing, which is consistent with enhanced OHM emission detected from the blueshifted outflowing molecular gas.}
	\label{fig:ohm_blue}
\end{figure}

\begin{figure}
	\centering
	\includegraphics[width=0.5\textwidth]{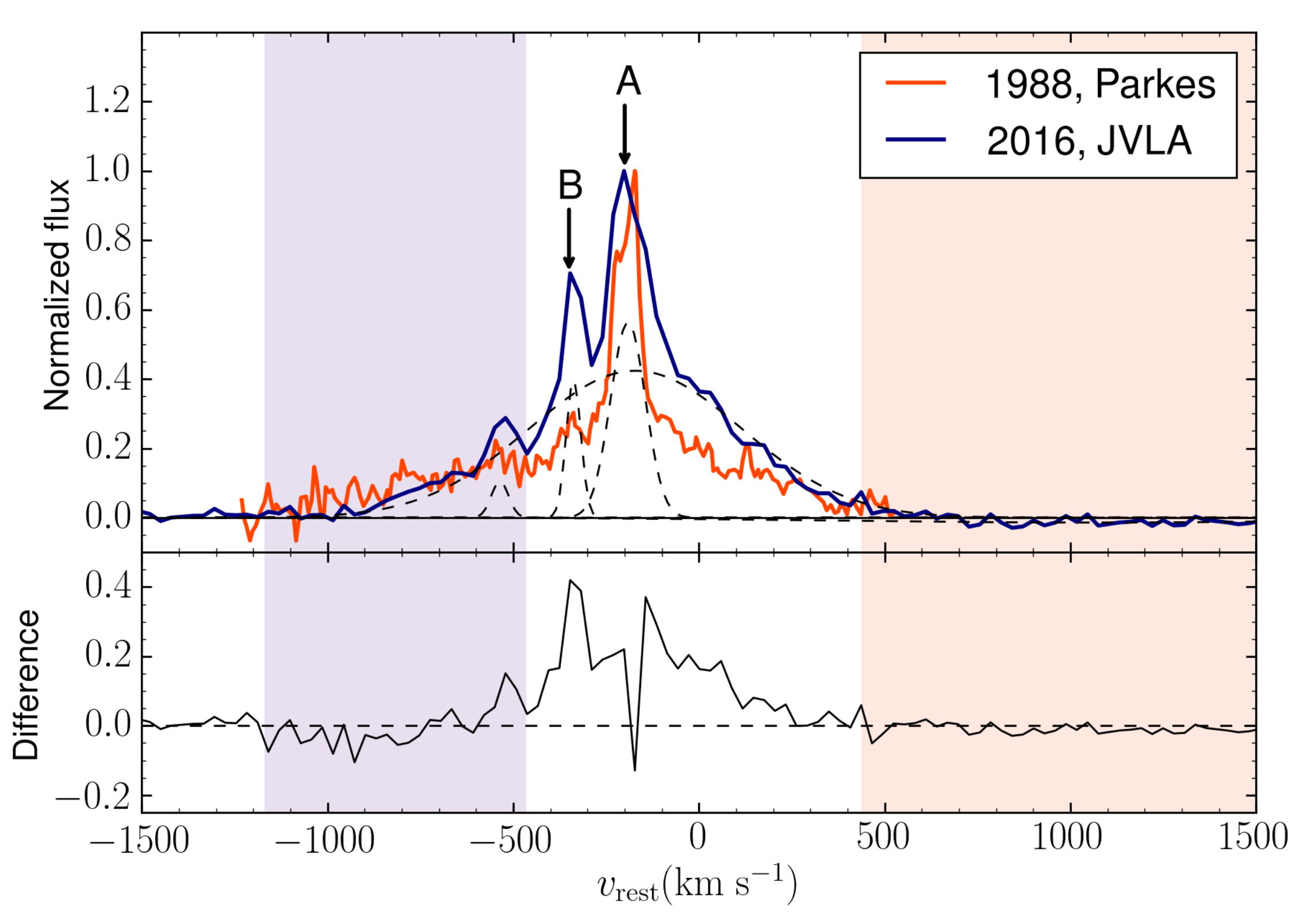}
	\caption{OHM spectra from IRAS 20100$-$4156 observed at different epochs, from 1988 to 2016. The spectrum from \citet{staveley1989} has not been continuum subtracted, due to the lack of an extended baseline i.e the normalization of the two spectra are uncertain. We here normalize to the same peak, to highlight changes in the spectral line profiles. The line profiles clearly show the presence of a varying narrow line component at $\sim -350$ \kms, and possibly one at $\sim -550$ \kms. These suggest the presence of compact masing sources with sizes of $\sim 10$ pc, which would be consistent with previous VLBI observations of OH megamasers. } 
	\label{fig:ohm_time}
\end{figure}
\begin{figure}
	\centering
	\includegraphics[width=0.5\textwidth]{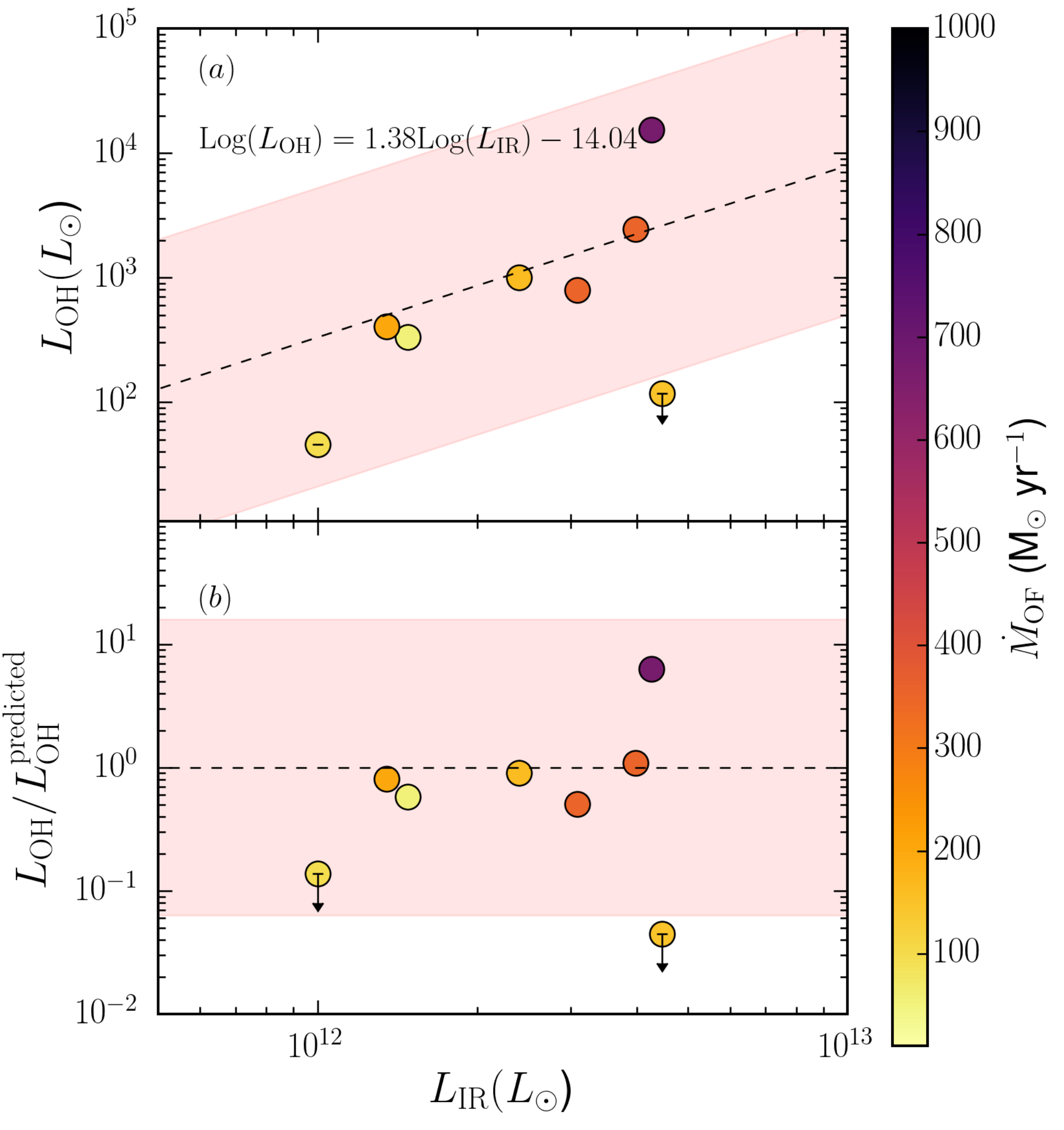}
	\caption{(a) The OHM luminosity as a function of the IR luminosity $L_{\rm IR}$. The $L_{\rm OH}-L_{\rm IR}$ correlation from \citet{yu2003} is shown as the black dashed line, and the shaded pink region shows the $\pm 1.2$ dex spread around the best-fit relation. All outflow sources that have been observed in OHM emission have been shown, compiled from \citet{yu2003,willett2011} (b) $L_{\rm OH}$, normalized by the predicted $L^{\rm predicted}_{\rm OH}$ from the $L_{\rm IR} - L_{\rm OH}$ relation. The color of the points represents the CO-inferred mass outflow rate $\dot M_{\rm OF}$. ULIRGs with higher mass outflow rates $\dot M_{\rm OF}$ have higher $L_{\rm OH}$.}
	\label{fig:ohm_mout}
\end{figure}

We report the first detection of OHM emission from IRAS 03158$+$4227 (\autoref{fig:iras03158_allspecs}) and a detection of time-variable OHM emission from IRAS 20100$-$4156, including significant emission from its high-velocity outflow wings (\autoref{fig:iras20100_allspecs}, \autoref{fig:iras20100infall}). The OHM emission in IRAS 20100$-$4156 was first reported by \citet{staveley1989}, although they could not confirm the high-velocity outflow wings due to limited bandwidth and non-linearities in the spectral baseline in their single-dish observations. More recent observations of IRAS 20100$-$4156 were conducted independently in 2015 using the Australia Compact Array (ATCA) and Australian Square Kilometre Array Pathfinder (ASKAP)'s Boolardy Engineering Test Array (BETA), where high-velocity wings were also detected, although with a lower significance \citep{hs2016}. These authors interpreted the high-velocity components as maser emission from molecular gas clumps rotating about a central SMBH ($M_{\rm BH} \sim 2.8\times 10^{9} M_{\odot}$, radius $r \sim 50$ pc), with the narrow-line components potentially associated with double nuclei, or with an outflow. 

Our observed OHM spectrum for IRAS 20100$-$4156 is consistent with the published ASKAP/BETA spectra, and has significantly better sensitivity. The OHM emission from IRAS 20100$-$4156 can be decomposed into a broad component, together with multiple narrow components (\autoref{fig:iras20100infall}). It shows a strikingly different velocity profile as compared to the CO($1-0$), with a systemic blueshift in the line core, enhanced emission from the blueshifted outflow wing, and tentative evidence of redshifted absorption in the OHM line profile (\autoref{fig:iras20100infall}). Despite these differences, the velocity extent of the detected wings agrees remarkably well with previous detections of the outflowing molecular gas - CO($1-0$) and OH 119 $\mu$m. This is excluding the velocity offset between the CO and OHM peak emission, which is not surprising given the extremely compact nature of OHM emission (e.g $\lesssim 0.2$ kpc; \citealt{pill2005}). The peak of the OHM emission could also be tracing the low-velocity outflow ($v_{\rm outflow} \sim 200$ \kms), which would be in line with the systematic blueshift of OH $65$ \mum$ $ absorption detected in ULIRGs \citep{ga2017}. 

We detect a spatial offset of $\sim 0.9 \pm 0.3$ kpc between the OHM emission (spatially unresolved) and the peak of the CO($1-0$) emission in IRAS 20100$-$4156, indicating a non-circumnuclear origin for the maser emission unlike previous OHM maps \citep[e.g.][]{pihlstrom2001, klockner2003, rovilos2003}, as shown in \autoref{fig:ohm_blue}. The OHM emission is co-spatial with the blue outflow wing as detected in CO, and at a small offset from the peak of the red outflow wing emission. While numerous previous observations have reported asymmetries in OHM emission, potentially due to outflows \citep[e.g.][]{baan1989b, baan1992}, this is the first direct comparison between spatially resolved powerful molecular outflows detected in CO and OHM that we are aware of. 
 
Spatially resolved OHM observations have revealed that the emission comprises two components: compact, high-gain saturated OHM emission on $\sim 1-10$ pc scales, and low-gain unsaturated diffuse emission over larger scales ($\sim 100$ pc; e.g. \citealt[][]{lonsdale1998,lonsdale2003}). The observed OHM spectrum for IRAS 20100$-$4156 is consistent with this picture if the broad pedestal traces the large-scale diffuse emission, and the narrow features traces dense embedded clumps of OHM emission. OHM emission is also predominantly from warm, dense molecular gas with $n_{\rm H_{2}} \sim 10^{4}$ cm$^{-3}$ and gas/dust temperatures of $\gtrsim 100$ K, allowing us to put the first constraints on the gas density and temperature in the outflowing molecular gas in IRAS 20100$-$4156. Such physical conditions, favorable for OH masing, are also associated with intense star formation activity \citep[see][for a review]{lo2005}, which is also reflected in the strong correlation between the occurrence of OHM emission and high dense molecular gas fractions (which trace actively star-forming molecular gas, \citealt{gao2004a, gao2004b}), traced by HCN \citep{darling2007, willett2011}. It is therefore interesting to explore whether the excess emission in the outflow wings is tracing ongoing star formation in outflowing molecular gas (e.g. \citealt{maiolino2017}). 

One of the more puzzling aspects of the OHM emission is the marked asymmetry between the red and blue outflow wing emission. In the velocity range of the red outflow wing, we instead tentatively detect redshifted absorption (\autoref{fig:iras20100infall}, B), which implies the presence of continuum emission behind the redshifted outflowing gas along the line of sight. One explanation is that this continuum emission is synchrotron emission, either from relativistic electrons generated by the forward shock associated with the molecular outflow \citep[see][]{ga2018}, or due to a localized starburst where the outflow interacts with the galaxy-wide ISM. This continuum emission can then act as the background for the OH 1.667 GHz redshifted absorption relative to the observer. If this scenario is correct, the detection of OH 1.667 GHz emission/absorption from outflowing gas in other ULIRGs will depend sensitively on the outflow opening angle and the relative orientation of the outflow axis to our line-of-sight. Alternatively, it is possible that there is a compact secondary AGN nucleus in IRAS 20100$-$4156 along the line of sight, and the receding outflow component is seen in OHM absorption against the continuum emission from this nucleus \citep[e.g.][]{ballo2004}. In order to differentiate between these hypotheses, observations with much higher spatial resolution are required. 

The multi-epoch observations of OHM emission from IRAS 20100$-$4156 also allow for identification of time-variable components, which in turn constrain the size-scales of the emission. In \autoref{fig:ohm_time}, we compare the normalized emission from \citet{staveley1989} to our observations. Given the limited bandwidth of the Parkes observations which do not cover any non-line continuum, we do not perform continuum subtraction and assume that the peak of the spectral line remains constant. This allows us to identify any relative changes in the line profile. We find a significant excess in OHM emission in 2016 as compared to 1988, with the largest change in component B (see \autoref{fig:ohm_time}), and with a smaller increase between $-500 $ \kms$ $ to $500$ \kms. This observed increase in emission from component B is consistent with that detected in \citet{hs2016}. The time-variation in OHM emission from the narrow line region implies an origin of a length scale $\lesssim 10$ pc, which is consistent with compact luminous masing clumps. 

We also test whether OHM emission is preferentially enhanced in molecular outflow hosts by compiling previous OHM observations for our sample of powerful molecular outflow sources (\autoref{fig:ohm_mout}). We compare them to the previously determined $L_{\rm OH} - L_{\rm IR}$ correlation \citep{darling2002a, darling2002b}. We find that stronger molecular outflows appear to be preferentially detected in sources with increased OHM luminosity, after controlling for the correlation with the IR luminosity. Outflow observations using OHM emission are also observationally cheaper than CO - the peak-to-line wing ratio in IRAS 20100$-$4156 is $\sim 7$ times higher than that for CO($1-0$). This translates to a relative reduction by a factor of $\sim 50$ in observing time between CO and OHM emission. All of these suggest that OHM activity is an exciting avenue in which to explore the properties of molecular outflows in a much larger sample of galaxies, and at higher redshifts (up to $z \sim 0.7$ with the VLA).

\section{Conclusions and future work}\label{sec:conclusions}
 
We have presented spatially resolved molecular gas observations of the two most powerful outflows in the local universe, and placed them in the context of other known molecular outflows in ULIRGs drawn from the same parent sample. 

\begin{itemize}

\item Our observations reveal extremely fast ($v_{\rm max} \gtrsim 1600$ \kms), spatially resolved, biconical molecular outflows in IRAS 20100$-$4156 and IRAS 03158+4227, and conservative mass outflow rates of $\sim 300-700 M_{\odot}$ yr$^{-1}$. Based on their observed gas masses, we infer gas depletion timescales of $\tau_{\rm OF}^{\rm dep} \sim 11$ and $\tau_{\rm OF}^{\rm dep} \sim 16$ Myr respectively. These are around $\sim 3$ times shorter than the gas depletion timescales due to star-formation, and demonstrate that extreme molecular outflows can starve the gas supply in galaxies on short timescales. 

\item We use multiple diagnostics to determine the AGN-fraction $f_{\rm AGN}$ in the two target sources. Based on mid-IR decomposition, X-ray observations, and SED fitting, we do not find a significant AGN fraction for IRAS 20100$-$4156, and a modest one for IRAS 03158+4227. 

\item We use mid-IR decomposition for an expanded sample of all published ULIRGs with molecular outflows observed using CO($1-0$), and test for correlations between them. We find that extreme molecular outflows can be hosted by AGN-dominated, starburst dominated, and composite galaxies i.e. $f_{\rm AGN}\sim 0.2 - 0.9$, and that the outflow energy and momenta are equally well correlated with AGN and starburst luminosities. While the outflow in IRAS 03158+4227 is consistent with being AGN-driven, we suggest that nuclear star formation may play an important role in driving the molecular outflow in IRAS 20100$-$4156. 

\end{itemize}

We also highlight some of the future observations that are necessary for understanding the impact of molecular outflows on their host galaxies, and for better characterizing the outflow mass and energetics. This includes observations of (1) spatially resolved CO($1-0$) in the molecular gas for a larger sample of ULIRGs to accurately obtain the outflow geometry and radius, (2) optically thin gas tracers, to accurately trace the outflow gas mass, and therefore the momentum and energy flux (3) future continuous monitoring of OH megamaser emission, providing very exciting avenues for future research. Observations such as these are well within the reach of current radio/mm/sub-mm interferometers (NOEMA, ALMA, and the VLA), and will allow us to build a comprehensive picture of the role of outflows in galaxy evolution. 

\acknowledgements

We thank the anonymous referee for their helpful comments which significantly improved the clarity of the manuscript. We are grateful to E. Nardini for helping us with the starburst and AGN templates for the mid-IR decomposition. We thank Prof. Lister Staveley-Smith for providing us with Parkes OH megamaser spectra for IRAS 20100$-$4156. We also thank H. Meusinger for providing us with optical data for IRAS 03158+4227. A.G thanks Cody Lamarche for many fruitful discussions. D.R. acknowledges support from the National Science Foundation under grant number AST-1614213 to Cornell University. E.GA is a Research Associate at the Harvard-Smithsonian Center for Astrophysics, and thanks the Spanish Ministerio de Econom\'{\i}a y Competitividad for support under project ESP2015-65597-C4-1-R, and thanks NASA grant ADAP NNX15AE56G. Basic research in IR astronomy at NRL is funded by the US-ONR; J.F. acknowledges support from NHSC/ JPL subcontracts 139807 and 1456609. CF acknowledges support from the European Union Horizon 2020 research and innovation programme under the Marie Sklodowska-Curie grant agreement No. 664931. J.A. gratefully acknowledges support from the Science and Technology Foundation (FCT, Portugal) through the research grant PTDC/FIS-AST/2194/2012 and UID/FIS/04434/2013. This paper makes use of the following ALMA data: ADS/JAO.ALMA\#2013.1.00659.S. ALMA is a partnership of ESO (representing its member states), NSF (USA) and NINS (Japan), together with NRC (Canada) and NSC and ASIAA (Taiwan) and KASI (Republic of Korea), in cooperation with the Republic of Chile. The Joint ALMA Observatory is operated by ESO, AUI/NRAO and NAOJ. The National Radio Astronomy Observatory is a facility of the National Science Foundation operated under cooperative agreement by Associated Universities, Inc. Funding for the Sloan Digital Sky Survey IV has been provided by the Alfred P. Sloan Foundation, the U.S. Department of Energy Office of Science, and the Participating Institutions. SDSS-IV acknowledges support and resources from the Center for High-Performance Computing at the University of Utah. The SDSS web site is www.sdss.org. Based on observations carried out under project ID W08E with the IRAM NOEMA Interferometer. IRAM is supported by INSU/CNRS (France), MPG (Germany) and IGN (Spain). This publication is based on data acquired with the Atacama Pathfinder Experiment (APEX) under programme ID 090.B-0404. APEX is a collaboration between the Max-Planck-Institut fur Radioastronomie, the European Southern Observatory, and the Onsala Space Observatory. This research has made use of the NASA/IPAC Extragalactic Database (NED) which is operated by the Jet Propulsion Laboratory, California Institute of Technology, under contract with the National Aeronautics and Space Administration.

\facilities{ALMA, EVLA, HST, APEX, IRAM, \emph{Herschel}, \emph{Spitzer}}
\software{CASA \citep{mcmullin2007}, GILDAS \citep{pety2005}, Astropy \citep{astropy2013}, CIGALE \citep{noll2009}, DS9, GAIA Skycat}

\appendix 

\begin{figure*}
	\includegraphics[width=0.5\textwidth]{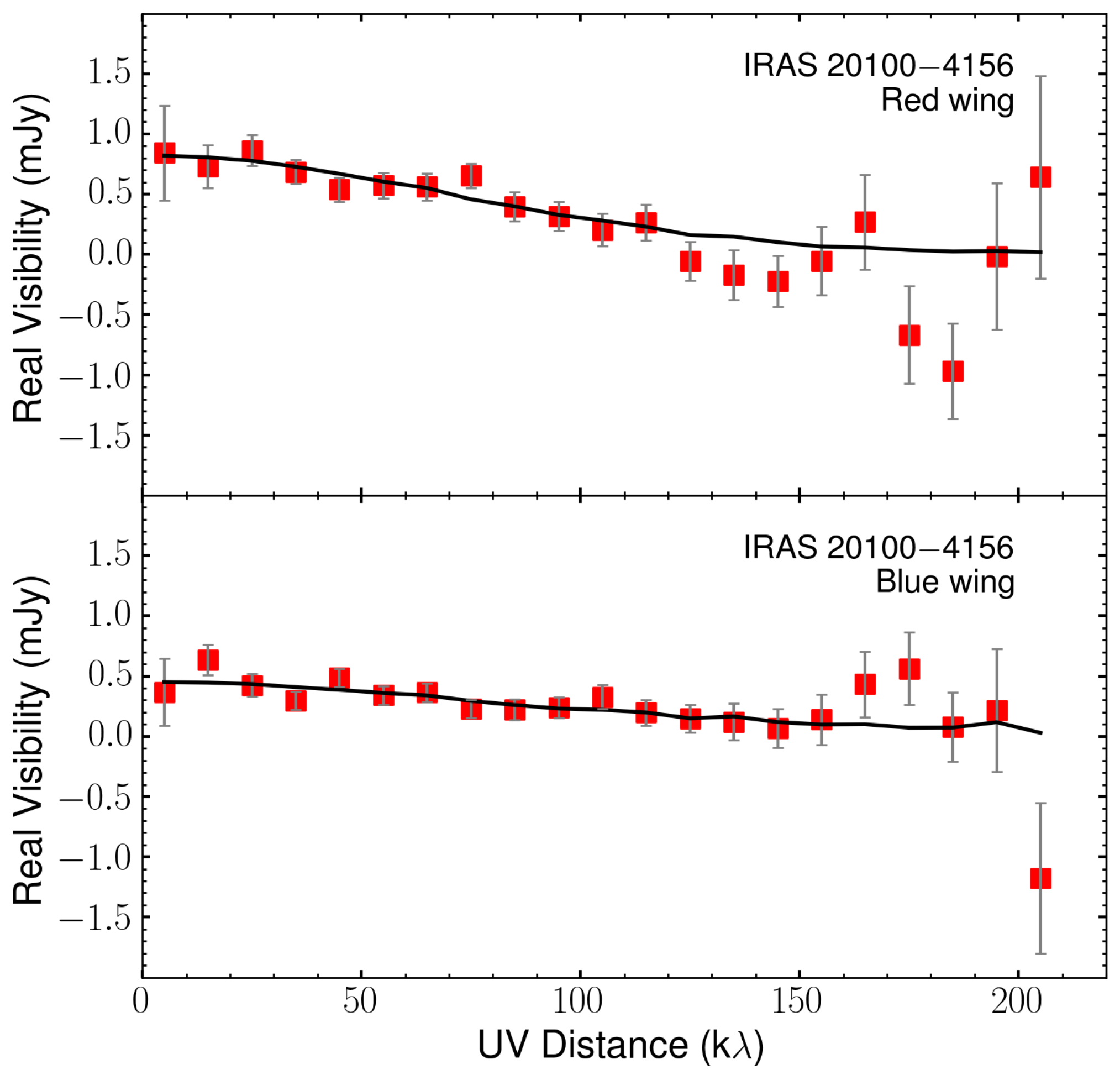}
	\includegraphics[width=0.5\textwidth]{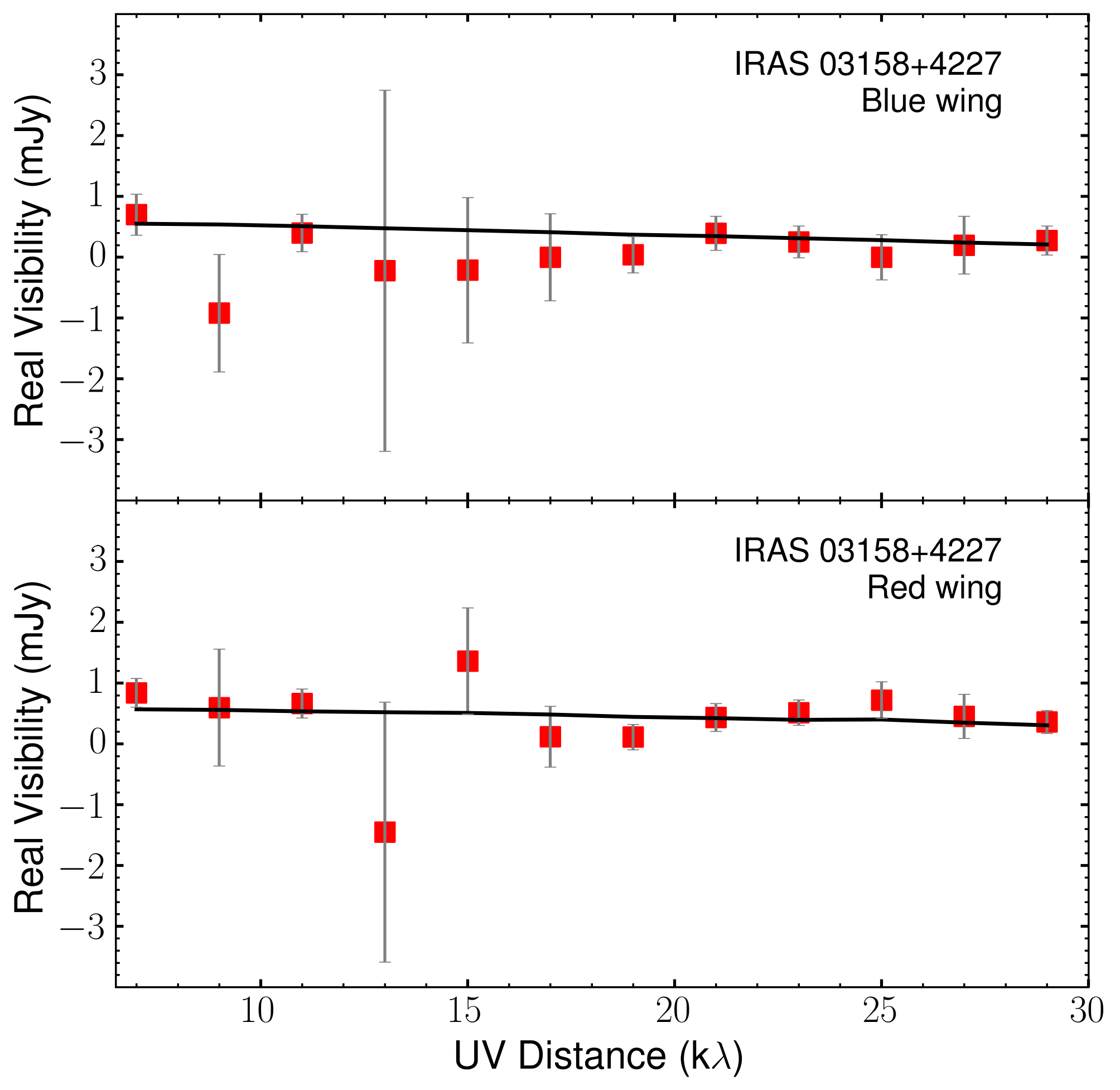}
	\caption{Results of using {\sc uvmodelfit} to model the visibilities for each of the red and blue outflow wings in IRAS 03158$+$4227 and IRAS 20100$-$4156, assuming that the outflow emission can be represented by a 2D Gaussian. }
	\label{fig:bothuvplots}
\end{figure*}

Here we present the details of model-fitting to the visibilities for the red- and blue-shifted outflow wings in IRAS 20100$-$4156 and IRAS 03158+4227. The fits to the \emph{uv} visibility data for the red and blue wings are described by six parameters: the integrated flux (I), offsets from the phase center ($\delta x$ and $\delta y$), the FWHM along the major axis ($a$), the axial ratio between the semi-minor and semi-major axes ($r$), and the position angle (PA). Note that the phase-centers were shifted to the centroids of the red and blue-shifted emission from the moment-0 maps before the fitting, resulting in the small offsets found in the model. The best $uv$-model fits are shown in \autoref{fig:bothuvplots}, and their parameters are listed in \autoref{tab:uvmodel}. We see evidence that a 2-D Gaussian does not fully reproduce the data for IRAS 20100$-$4156 on the longest baselines, indicating that the red and blue outflow wings are individually marginally resolved in our observations and that they deviate from elliptical Gaussians on the smallest scales (\autoref{fig:bothuvplots}). For IRAS 03158+4227, the visibilities suffer from greater uncertainties on shorter baselines (i.e. larger spatial scales) due to a relatively sparse $uv$-coverage on those scales. 

\begin{table*}
\setlength{\tabcolsep}{3pt}
\renewcommand{\arraystretch}{1.1}
\begin{center}
\caption{Description of \textsc{uvmodelfit} results for red and blue-shifted emission.}
\begin{tabular}{lllll}
\tableline
\tableline
Parameter 		&
\multicolumn{2}{c}{IRAS 20100$-$4156} 	&
\multicolumn{2}{c}{IRAS 03158+4227}  \\ 
				&  Red wing 	& Blue wing 	& Red wing  	& Blue wing	 \\ \toprule
RA\tablenotemark{a} & $20^h13^m29.577s $  & $20^h13^m29.536^s$ &$03^h19^m11.896^s $ & $03^h19^m12.003^s$ \\
Dec\tablenotemark{a} & $-41^d47^m35.550^s$  & $-41^d47^m34.997^s$ &$+42^d38^m25.374^s$ & $+42^d38^m24.285^s$ \\
I ($ 10^{-4}$Jy)				& $4.3 \pm 0.3 $ &$ 8.2 \pm 0.4 $&$ 5.9 \pm 1.6 $	& $ 5.9 \pm 2.8 $ \\ 
$\delta x$ (arcsec) 		& $0.07 \pm 0.03 $ & $-0.03 \pm 0.03$ & $0.1 \pm 0.4$ & $0.03 \pm 0.8 $\\ 
$\delta y$ (arcsec)		& $0.00 \pm 0.03 $&$ 0.05 \pm 0.03 $& $-0.01 \pm 0.40 $ & $0.02 \pm 0.80$ \\
$a$(arcsec)	& $0.8 \pm 0.1 $ & $1.3 \pm 0.1 $&$ 3.3 \pm 1.3$ & $3.8 \pm 2.6$ \\
$r$				& $1.0 \pm 0.2  $&$ 0.6 \pm 0.1$ &$ 0.8 \pm 0.5 $& $1.0 \pm 0.9$ \\
PA ($\degree$)				& $74 \pm 57$& $-28 \pm 8 $ & $ -31 \pm 82 $&$ -52 \pm 57$ \\
\tableline
\end{tabular}
\tablenotetext{a}{The assumed phase centers for the red and blue-shifted emission, obtained from the centroid of the moment-0 maps for each.}

\label{tab:uvmodel}
\end{center}
\end{table*}

%\bibliography{nat}

\end{document}